\documentstyle[emulateapj]{article}

\lefthead{Furusawa et al.}
\righthead{New Improved Photometric Redshifts of Galaxies in the HDF}

\newcommand{\Deriv}[2]{\frac{\partial}{\partial #2}\,#1}

\newcommand{\Deltait}{{\it \Delta}}
\newcommand{\photoz}{$z_{\rm photo}$}
\newcommand{\specz}{$z_{\rm spec}$}
\newcommand{\chisq}{$\chi^2$}

\begin{document}

\title{New Improved Photometric Redshifts of Galaxies in the HDF}

\author{Hisanori Furusawa,
Kazuhiro Shimasaku\altaffilmark{1},
Mamoru Doi\altaffilmark{1},
and Sadanori Okamura\altaffilmark{1}}

\affil{Department of Astronomy, 
School of Science, University of Tokyo,\\ Bunkyo-ku, Tokyo 113-0033, Japan}

\altaffiltext{1}{Research Center for the Early Universe,
University of Tokyo, Japan}

\begin{abstract}
We report new improved photometric redshifts of 1048 galaxies 
in the Hubble Deep Field (HDF). 
A standard $\chi^2$ minimizing method is applied to seven-color
{\it UBVIJHK} photometry by Fern\'{a}ndez-Soto, Lanzetta, \& Yahil
 (1999).  
We use 187 template SEDs representing a wide variety of 
morphology and age of observed galaxies based on a population 
synthesis model by Kodama \& Arimoto (1997). We introduce two new
recipes. First, the amount of the internal absorption is changed
as a free parameter in the range of $E(B-V)=0.0$ to $0.5$ with an 
interval of $0.1$.  Second, the absorption due to intergalactic HI clouds 
is also changed by a factor of 0.5, 1.0, and 1.5 around the
opacity given by Madau (1995). The total number of template SEDs
is thus $187\times 6\times 3=3,366$, except for the redshift grid.
The dispersion $\sigma_z$ of our photometric redshifts with respect 
to spectroscopic redshifts is $\sigma_z=0.08$ and $0.24$ for $z<2$ and
$z>2$, respectively, which are smaller than the corresponding values
($\sigma_z=0.09$ and $0.45$) by Fern\'{a}ndez-Soto et al. 
Improvement is significant, especially in $z>2$.  This is due to 
smaller systematic errors which are largely reduced mainly by 
including three opacities due to intergalactic HI clouds.
We discuss redshift distribution $N(z)$ and cosmic star formation rate 
based on our new photometric redshifts.
\end{abstract}


%%%%%%%%%%%%%%%%%%%%%%%%%%%%%%%%%%%%%%%%%%%%%%%%%%%%%%%%%%%%%%%%%%%%%%%%%
%                            Section 1                                  %
%%%%%%%%%%%%%%%%%%%%%%%%%%%%%%%%%%%%%%%%%%%%%%%%%%%%%%%%%%%%%%%%%%%%%%%%%

\section{INTRODUCTION}\label{sec:intro}
Large redshift surveys of galaxies have played a quite important role
for understanding the formation and evolution of galaxies.  In 1990's,
several new deep redshift surveys using large optical telescopes
were carried out: e.g., 
Canada-France Redshift Survey (CFRS; Lilly et al. 1995), Hawaii
Deep Field (Cowie et al. 1996), Autofib/LDSS survey (Ellis 
et al. 1996), the Canadian Network for Observational 
Cosmology (CNOC1,2 surveys; Yee et al. 1996 and 
Carlberg et al. 1998). 
Some of those surveys reach as deep as $z\sim 1$, and we have obtained
some information about the formation and evolution of galaxies up to
$z\sim 1$.  However, little has been known about the properties of $z>1$
galaxies from such redshift surveys due to the limited depth of
spectroscopic observations.
Even if one uses LRIS with a Keck telescope, the limiting magnitude
of spectroscopy is about R$_{\rm lim}\simeq 25$, while the limiting
magnitude of photometry with Hubble Space Telescope ({\it HST}) WFPC2 is $R_{\rm lim} \simeq 28-29$. 

There is, however, so called `photometric redshift technique', which
enables us to measure redshifts of faint galaxies.
Photometric redshift technique was originally suggested by Baum et
al. (1963) to determine redshifts of some clusters of galaxies based on
multi-band photometry using mean spectra of their member galaxies.
There were, however, only a few applications of the technique 
(e.g., Koo 1985; Loh \& Spillar 1986) before it revived recently to study distant galaxies,
in particular, faint galaxies found in the Hubble Deep Field (HDF;
Williams et al. 1996) (Connolly et al. 1995; Lanzetta, Yahil, \&
Fern\'{a}ndez-Soto 1996, 1998; Gwyn \& Hartwick 1996; Mobasher et
al. 1996; Sawicki, Lin, \& Yee 1997; Connolly et al. 1997).
Photometric redshift technique is the only method available at present
to infer the redshifts of galaxies beyond the spectroscopic limit.

In this work we improve the accuracy of photometric redshifts of
galaxies in the HDF by using a set of template spectral energy
distribution (SED). We treat the internal absorption of galaxies and
the extinction due to intergalactic HI clouds more carefully than 
previous studies.
We adopt a technique based on a standard $\chi^2$ 
minimizing method with a template set which consists of simulated SEDs. 
We apply it to a photometry catalog of galaxies in the HDF
(Fern\'{a}ndez-Soto, Lanzetta, \& Yahil 1999), which is 
the deepest catalog available with some spectroscopic redshifts.
 
The detailed method, adopted parameters of simulated SEDs, and the
comparison between our photometric redshifts and spectroscopic redshifts 
are described in \S~2. We show in \S~3 the comparison of our
improved photometric redshifts with previous studies and the validity of
our result.
We present our improved photometric redshift for the HDF galaxies in 
\S~4, where we also discuss the properties of the HDF galaxies such 
as redshift distribution and cosmic star formation rate.  Conclusions
are given in \S~5.

%%%%%%%%%%%%%%%%%%%%%%%%%%%%%%%%%%%%%%%%%%%%%%%%%%%%%%%%%%%%%%%%%%%%%%%%%
%                            Section 2                                  %
%%%%%%%%%%%%%%%%%%%%%%%%%%%%%%%%%%%%%%%%%%%%%%%%%%%%%%%%%%%%%%%%%%%%%%%%%

\section{METHOD}

We may classify photometric redshift techniques into two categories.
One is the training-set method (Connolly et al. 1997, Wang et al. 1999) 
and the other is the $\chi^2$ minimizing method (Gwyn \& Hartwick 1996;
; Mobasher et al. 1996, 1999; Sawicki et al. 1997; Giallongo et al. 1998).

The training-set method first makes an empirical formula by fitting
polynominals to the photometry data of galaxies with known redshifts.
It then applies the same formula to photometry data of new galaxy samples to
determine their photometric redshifts.
This method is superior to the other in the point that the more galaxies we
observe for the training set, the more accurate photometric redshifts %
we obtain.
Therefore, Yee (1998) called it the empirical training-set method.
For low-{\it z} galaxies, the training-set method seems to work more robust
than the \chisq\ method, because it is based on real observed 
galaxy spectra and because the empirical formula can be determined
accurately using a large number of spectroscopic samples (Connolly et al. 1997).
%change
Recently, comparably accurate redshifts are obtained by a simpler
training-set method which adopts binning of galaxies according to
observed colors (Wang, Bahcall \& Turner 1998; Wang, Turner, \& Bahcall 1999).  Csabai, Connolly, \& Szalay (1999) also succeeded in obtaining accurate photometric redshifts by combining the training-set method and the \chisq\ method. 
However, it is difficult to make a good training set for high redshift %
galaxies, since the number of galaxies with spectroscopic redshifts is %
still small. 
Also we should be cautious in applying the empirical formula obtained %
at low redshifts to high redshift galaxies, since significant evolution %
is expected in their SEDs.

Hence, in this study, we use the \chisq\ minimizing method which can include
spectral evolution of galaxies directly.
This method searches for the best-fit SED and the redshift of a galaxy by
comparing the observed SED of the galaxy with templates prepared in advance.
In general, templates are made either from observed spectra of nearby 
galaxies or from simulated spectra based on a population synthesis model.
Coleman, Wu, \& Weedman (1980; hereafter CWW) and Kinney et al. (1993, 1996) are widely used sources of observed spectra for the HDF
galaxies.
Lanzetta et al. (1996, 1998) and Fern\'{a}ndez-Soto et al. (1999) used
CWW's and Kinney et al.'s spectra.
However, the use of the observed SEDs of nearby galaxies may make it
difficult to directly take the effects of galaxy evolution into account.
Accordingly, in this paper we use simulated spectra based on a recent
population synthesis model. 
Gwyn \& Hartwick (1996) also used simulated spectra based on Bruzual \& 
Charlot (1993). Sawicki et al. (1997) made their templates based
on both observed spectra (CWW) and simulated spectra by Bruzual \&
Charlot (1996; hereafter BC96).

\subsection{Template SEDs}\label{sec:template}

We use a stellar population synthesis model by Kodama \&
Arimoto (1997; hereafter KA97) to make template SEDs. 
KA97 includes the stellar evolutionary tracks after the asymptotic 
giant branch in addition to the conventional tracks.
Hence the model predicts UV flux of galaxies reasonably well, or at least, 
as good as any previous models.
KA97 was successfully used to obtain photometric redshifts of %
low redshift galaxies (Kodama, Bell, \& Bower 1999).  Our work is the first %
to use KA97 to obtain photometric redshifts of high-$z$ galaxies.  

The template SEDs consist of the spectra of pure disks, %
pure bulges, and composites made by interpolating the two
as shown in Table~\ref{table:sedparam}.
The parameters for the SEDs are the power-law index $x^{\rm IMF}$ of the
initial mass function (IMF), the time scale of star formation $\tau_{\rm SF}$,
the time scale of gas infall from a galactic halo into a disk $\tau_{\rm
infall}$, and the time when the galactic wind blows $t_{\rm GW}$.  The
star formation rate of a galaxy is set to be zero after $t_{\rm GW}$.
For disks, we adopt $x^{\rm IMF}=1.35$, $\tau_{\rm SF}=5$Gyr, $\tau_{\rm
infall}=5$Gyr, and $t_{\rm GW}=20$Gyr (i.e., longer than the present age
of the universe), which are close to the values estimated for the disk
of our Galaxy.  We adopt from KA97 $x^{\rm IMF}=1.10$, $\tau_{\rm
SF}=0.1$Gyr, $\tau_{\rm infall}=0.1$Gyr, and $t_{\rm GW}=0.353$Gyr for bulges, which are known to reproduce the average color of ellipticals in clusters of galaxies.
We make intermediate SED types by combining a disk component and a bulge 
component with the same age.  
%change
The ratio of the bulge luminosity to the total luminosity in the {\it B} band, which we define as B/T, is changed from 0.1 to 0.9 with an interval of 0.1.  Pure disk SEDs correspond to young or active star-forming galaxies, and pure bulge SEDs correspond to elliptical galaxies.  
We also prepare very blue SEDs of age$<$1Gyr, corresponding to blue
star-forming galaxies reported by recent deep surveys.  In total, our basic template set consists of 187 SEDs.

%\placetable{table:sedparam}

\subsection{The Absorption Effects}\label{sec:effects}

To simulate the observed SEDs, we take into account two absorption 
effects. One is the internal absorption due to dust in each galaxy.
More light is scattered and absorbed by dust grains in shorter wavelengths.
Therefore, internal absorption is one of the key parameters 
for photometric redshift technique which is sensitive to wavelength
dependent effects.  
For simplicity, we assume that all galaxies have the same extinction
curve, although we change the absolute amount of extinction, indicated by 
$E(B-V)$, as a free parameter.

Typical features of extinction curves of nearby galaxies including our Galaxy 
are as follows:
(1) The amount of absorption basically increases from the infra-red
to the ultraviolet.
(2) Some extinction curves have a bump at 2175\AA, which is usually
explained mainly by graphite.  The existence of the bump is reported 
for our Galaxy (Seaton 1979; Cardelli et al. 1989), LMC (Fitzpatrick 1985), and M31 (Bianchi et al. 1996). 
However, no bump is reported in the extinction curve of SMC (Pr\'{e}vot
et al. 1984; Bouchet et al. 1985).  Gordon, Calzetti, \& Witt (1997) showed that
whether the bump is present or not cannot be explained by scattering or
geometrical effects, and that galaxies with high star formation activity
tend to have no bump in their extinction curves.

There are some studies in which an analytical formula was fitted to
observed extinction curves (Cardelli et al. 1989; Fitzpatrick 1986,1998).
Calzetti (1997a) proposed an analytical formula for nearby star-forming
galaxies, including the effect of scattering through star-forming 
regions.  He inferred that his formula holds for distant young galaxies because these galaxies are probably similar to nearby star-forming galaxies.

We calculated photometric redshifts using the Milky-Way-like %
extinction curve by Cardelli, Clayton, \& Mathis (1989) and the SMC-like
curve by Calzetti (1997b). As a result, Calzetti's curve gave better
results than Cardelli et al.'s. 
Therefore, we have decided to adopt the curve by Calzetti (1997b),
\begin{eqnarray}
 F_{obs}(\lambda ) = F_0(\lambda ) 10^{-0.4\,{E(B-V)}\,k(\lambda )},
\end{eqnarray}
where $F_{obs}(\lambda )$ is the observed flux and $F_0(\lambda )$ 
is the intrinsic flux without extinction.  If we normalize the absorption
strength by $k({B})-k({V})=1$, 
$k(\lambda )$ is expressed as:
\begin{eqnarray}
 k(\lambda) &=& 2.656(-2.156+1.509/\lambda -0.198/\lambda^2+0.011/\lambda^3)\nonumber\\
 & & {} +4.88\hspace{1.5cm} \mbox{for}\ 0.12\,{\rm\mu m} \le \lambda\le 0.63\,{\rm\mu m}\nonumber\\
 &=& {} \left[(1.86-0.48/\lambda )/\lambda -0.1\right] /\lambda +1.73\nonumber\\
 & & {}\hspace{2.5cm} \mbox{for}\ 0.63\,{\rm\mu m} \le \lambda \le 1.0\,{\rm\mu m}
\end{eqnarray}
It is not clear whether or not most of distant galaxies have a SMC-like
extinction curve because of the ambiguity of the UV flux.

%\placefigure{fig:dustcurve}
%\placefigure{fig:dusttest}

\begin{figure}
\epsscale{0.6}
\plotone{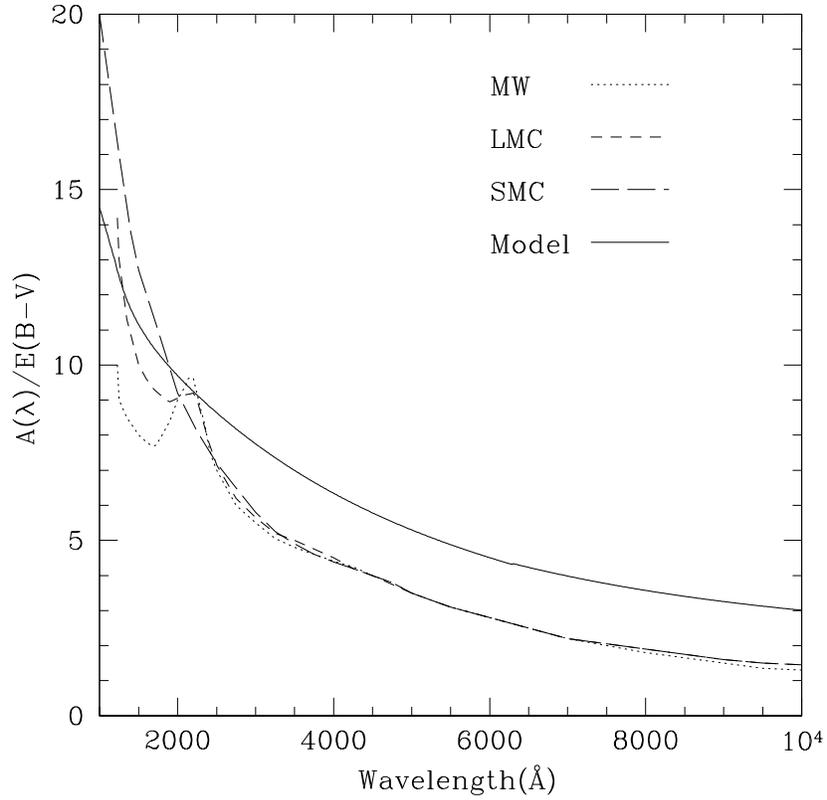}
\caption[fig1.ps]{The extinction curves of typical galaxies. The vertical axis
 means the absorption strength A($\lambda$), which is
 normalized by $E(B-V)=1$.  The curve given in Calzetti (1997b) is
 indicated as the solid line labeled `Model'. It is noted that Milky Way and LMC have the 2175\AA\ bump while SMC and Calzetti's curves do not.}
\label{fig:dustcurve}
\end{figure}

\begin{figure}
\epsscale{0.6}
\plotone{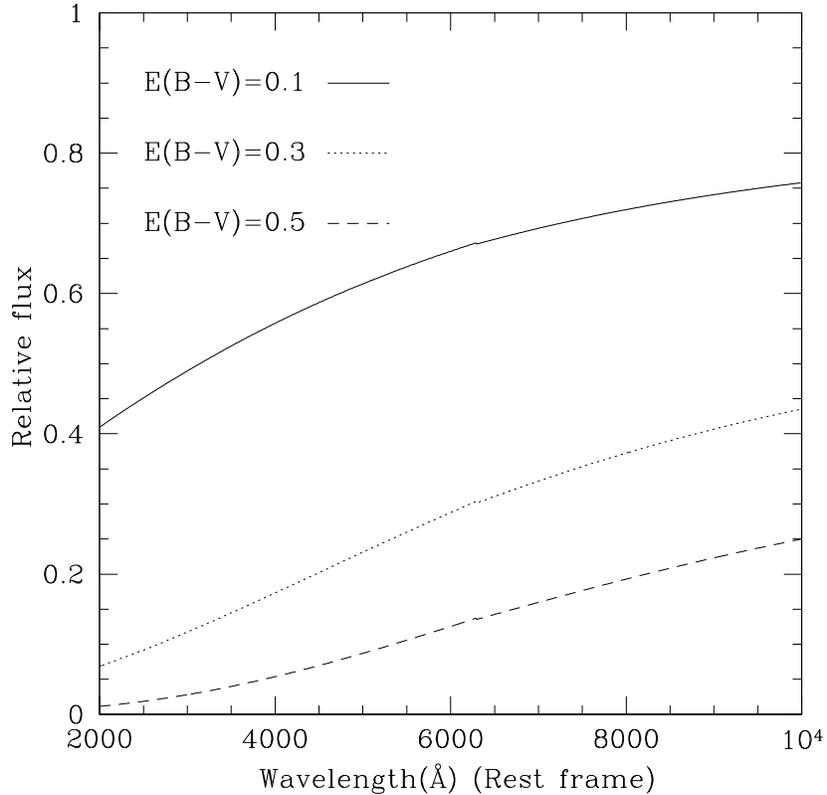}
\caption[fig2.ps]{The relative flux after absorption by interstellar dust in a galaxy. We assume Calzetti's extinction curve.}
\label{fig:dusttest}
\end{figure}

Figure~\ref{fig:dustcurve} shows the observed extinction curves and the
model curve calculated by Calzetti (1997b).  The curves are normalized so
that $E(B-V)$ is equal to $1$.
We make the template SEDs using the Calzetti's extinction curve with
$E(B-V)=0.0$ (dustless), 0.1, 0.2, 0.3, 0.4, and 0.5.  
Figure~\ref{fig:dusttest} shows the ratio of the flux after extinction with respect to the intrinsic flux as a function of wavelength for three values of $E(B-V)$.

\begin{figure}
\epsscale{0.6}
\plotone{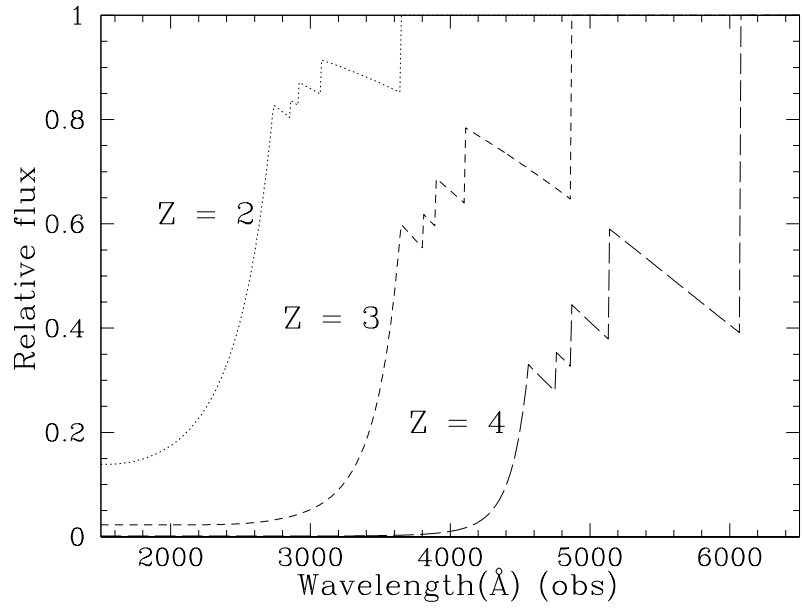}
\caption[fig3.ps]{The relative flux after absorption by intergalactic HI clouds 
for a galaxy at three redshifts.}
\label{fig:lymantest}
\end{figure}

\begin{figure}
\epsscale{0.6}
\plotone{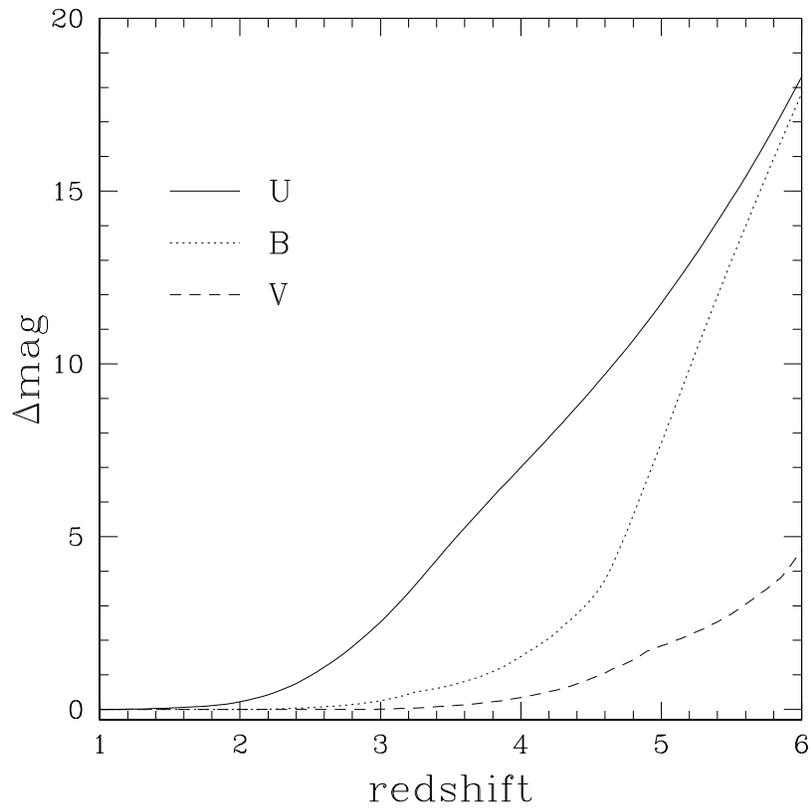}
\caption[fig4.ps]{The absorption strength in magnitude due to intergalactic HI
 clouds in case of observing a galaxy at $z=3.5$
 with broad band filters.  The {\it U}, {\it B}, and {\it V} bands here 
means F300W, F450W, and F606W filters of the {\it HST}, respectively.}
\label{fig:lymanmag}
\end{figure}

We also include the effect of absorption due to the intergalactic HI 
clouds.
On the assumption that the {\it b}-value, which is the indicator of the
velocity dispersion of the clouds, is 35km/s, Madau (1995) computed the
median optical depth of the line blanketing as 
\begin{eqnarray}
\tau_{\it eff}^{\it line} = \sum_{j=1,4} A_j\,
\left(\frac{\lambda_{\it obs}}{\lambda_j}\right)^{3.46},\label{eq:tauline}
\end{eqnarray}
where $j=1,2,3,$ and $4$ correspond to Ly-$\alpha$, Ly-$\beta$,
Ly-$\gamma$, and Ly-$\delta$ absorptions, respectively, 
$\lambda_j$=(1216, 1026, 973, 950\AA), and 
$A_j=(3.6\times 10^{-3}, 1.7\times 10^{-3}, 1.2\times 10^{-3}, 9.3\times 10^{-4})$.
The amount of absorption of lines at shorter wavelengths is negligible 
compared with these Ly-$\alpha$ to Ly-$\delta$ lines and Lyman
continuum.  Madau (1995) also showed that the continuum absorption 
below 912\AA\ including Lyman limit systems is expressed with a 5\%
accuracy by:
\begin{eqnarray}
\tau_{\it eff}^{\it cont} & \simeq & 0.25 x_c^3(x_{em}^{0.46}-x_c^{0.46}) +9.4 x_c^{1.5}(x_{em}^{0.18}-x_c^{0.18})\nonumber\\
{} & {} & {} -0.7 x_c^3(x_c^{-1.32}-x_{em}^{-1.32})\nonumber\\
{} & {} & {} -0.023 (x_{em}^{1.68}-x_c^{1.68}),\label{eq:taucont}
\end{eqnarray}
where $x_c = 1+z_c = \frac{\lambda_{\it obs}}{\lambda_L}$ and $x_{em} =
1+z_{em}$. 
Here, $\lambda_{\it obs}$ is the wavelength of the
observer's frame and $\lambda_L$=912\AA\ and $z_{em}$ is a redshift of an
emitting source. The total opacity of intergalactic 
HI clouds, $\tau_{\it eff}$, is expressed by the sum of
the opacity due to line absorption $\tau_{\it eff}^{\it line}$ and
that due to continuum absorption
$\tau_{\it eff}^{\it cont}$:
\begin{eqnarray}
\tau_{\it eff}=\tau_{\it eff}^{\it line}+\tau_{\it eff}^{\it cont}.\label{eq:taueff}
\end{eqnarray}
The relative flux due to intergalactic HI absorption is shown in Figure~%
\ref{fig:lymantest}.  We also estimate the amount of the HI absorption 
integrated within standard broad-band systems as shown in Figure~\ref{fig:lymanmag}. 

%\placefigure{fig:lymantest}
%\placefigure{fig:lymanmag}

We include a variation in opacity as follows. 
In addition to Madau's original median opacity ($\tau_{\it eff}$), 
we also use the opacity of 
$0.5\tau_{\it eff}(\equiv\tau^-_{\it eff})$ and 
$1.5\tau_{\it eff}(\equiv\tau^+_{\it eff})$.
%change
This is because we should take into account a variation of Lyman
absorptions statistically.  Here, $\tau^+_{\it eff}$ and $\tau^-_{\it
eff}$ are within $\pm 1\sigma$ of the opacity derived by Madau (1995).  
As we will see later, using three different values around the median 
value turns out to be more preferable to using just the median opacity.

\subsection{\chisq -Fitting and Constraints}\label{sec:constraints}
The fitting procedure to obtain reduced \chisq\ is summarized as follows.
We prepare a set of original 187 SEDs based on KA97,  
which cover a wide range of the spectral-type of observed galaxies. 
Then, we incorporate the absorption effects due to the internal dust and 
the intergalactic HI clouds (\S~\ref{sec:effects}).
Next, we make redshifted SEDs from $z=0.05$ to $z=8.0$ with an interval
of 0.05.  Finally, each SED at a given redshift is convolved by the band 
responses used for the observation.  We compute the flux of template 
galaxies observed in the $i$-th band by
\begin{eqnarray}
T_i &=& \frac{\int f_{z,\lambda}^{\it template} R^i_\lambda\,d\lambda}
             {\int R^i_\lambda\,d\lambda},\label{eq:convolv}
\end{eqnarray}
where $f_{z,\lambda}^{\it template}d\lambda$ is the flux
received in $[\lambda,\lambda+d\lambda]$ and $R^i_\lambda$ is the band 
response function of the $i$-th band.

We compare the convolved flux $T_{i}$ with the observed flux $F_{i}$ in
each band and calculate $\chi^2$ by:
\begin{eqnarray}
 \chi^2 &=& \sum_{i}^{N_{band}}\frac{(F_i - \alpha\,T_i)^2}
                                     {\sigma_i^2},\label{eq:chi2}
\end{eqnarray}
%change
where $N_{band}$ is the number of bands used for the observation,
$\sigma_i$ is an observational error in the $i$-th band.
If flux in the $i$-th band is not detected, $F_i$ is set to be zero.
$\alpha$ denotes the normalization factor which minimizes the 
reduced \chisq.  For a given set of $T_i$, the value of $\alpha$ which minimizes \chisq\ is
calculated by:
\begin{eqnarray}
\Deriv{\chi^2}{\alpha} &=& \Deriv{\sum_{i}^{N_{band}} \frac{(F_i - \alpha\,T_i)^2}{\sigma_i^2}}{\alpha} = 0,\label{eq:calc_alpha}\nonumber\\
 \alpha &=& \frac{\sum_{i}^{N_{band}} \frac{F_i T_i}{\sigma_i^2}}{\sum_{i}^{N_{band}} \frac{T_i^2}{\sigma_i^2}}.\label{eq:alpha}
\end{eqnarray}
Then, we find the best-fit SED which gives `the minimum reduced \chisq '
out of all the templates, and the redshift of the best-fit SED is
adopted as the photometric redshift of the galaxy. 

The most important point to get reliable photometric redshifts is the
validity of the template SEDs.  However, even with an ideal set of
templates, large estimation errors in $z_{\rm photo}$ caused by, for
example, small signal-to-noise ratios (S/N) of photometry are inevitable
in some cases. We should carefully check the results obtained by our
simple fitting method.
In this work, we impose two constraints in order to get robust and 
accurate estimations.
First, we reject the best-fit SEDs that give too bright or too faint
absolute magnitudes.  We then remove from our sample the best-fit SEDs
which have `minimum reduced \chisq ' larger than some threshold.  
It is found that the photometry data of too low S/N tend to give a large
value of `minimum reduced \chisq '.

%%%%%%%%%%%%%%%%%%%%%%%%%%%%%%%%%%%%%%%%%%%%%%%%%%%%%%%%%%%%%%%%%%%%%%%%%
%                            Section 3                                  %
%%%%%%%%%%%%%%%%%%%%%%%%%%%%%%%%%%%%%%%%%%%%%%%%%%%%%%%%%%%%%%%%%%%%%%%%%

\section{VERIFICATION OF THE METHOD}\label{sec:check}

\subsection{The Spectroscopic Sample}\label{sec:data}

We apply our method to galaxies in the HDF, %
where the deepest images of field galaxies were taken.  
We use the photometric catalog by Fern\'{a}ndez-Soto  et al. (1999), 
which is based on seven-band photometry, i.e., {\it UBVI} with the 
{\it HST} (Williams et al. 1996) and {\it JHK} taken with the 4m Mayall telescope at the Kitt Peak National Observatory (Dickinson et al. 1999). This catalog also gives one of the most recent estimates of photometric
redshifts (hereafter, \photoz ) of the HDF galaxies.
This catalog includes total magnitude and photometric errors for
galaxies of $I_{\rm AB}<28$ with a detection threshold of 
$\mu_{I_{\rm AB}}=26.1$ [mag arcsec$^{-2}$], though the magnitude limit
is $I_{\rm AB}=26$ in the outskirts of the field.
We summarize the limiting magnitudes in Table~\ref{table:obs_hdf}.
The number of galaxies in the catalog is 1067 in total, and 946 
galaxies are in the good S/N region, while 121 galaxies are in 
the outskirts. Out of them, 108 galaxies have spectroscopic
redshifts (hereafter, \specz ) (Cohen et al. 1996;  Steidel et al. 1996; 
Lowenthal et al. 1997; Zepf, Moustakas, \& Davis 1997; Cohen et
al. 2000), which we hereafter call the spectroscopic sample.

%\placetable{table:obs_hdf}

\subsection{Photometric Redshifts and SED Parameters}\label{sec:photoz}
\begin{figure}
\epsscale{0.8}
\plotone{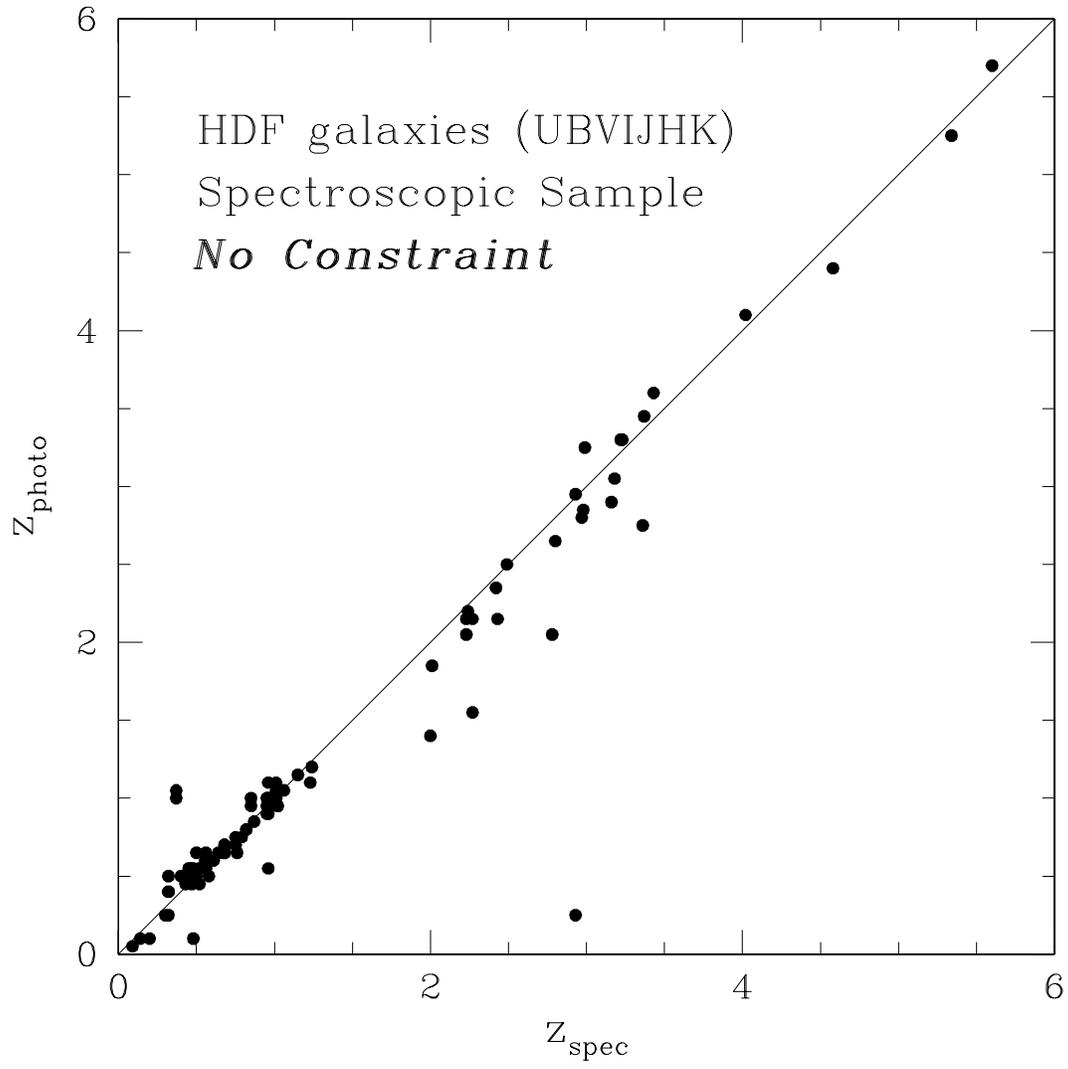}
\caption[fig5.ps]{Our \photoz\ versus \specz\ for the spectroscopic sample.   
The solid line is the equality line.}
\label{fig:photoz}
\end{figure}

We first measure \photoz\ for the spectroscopic sample,
using {\it UBVIJHK} seven-band magnitudes by the standard \chisq\
fitting of the template SEDs.  Figure~\ref{fig:photoz} shows 
the correlation diagram between our \photoz\ and \specz .
We cannot find the best-fit SED and fail to measure \photoz\ for two 
galaxies, and one galaxy at $z\simeq 3$ has a very large estimation
error, which is called the catastrophic error by Sawicki et al. (1997).
As seen in Fig~\ref{fig:photoz}, the redshifts of four galaxies at 
\specz $\simeq 2-3$ are underestimated by $\Deltait z\sim 0.5$.
A similar trend was seen in Fern\'{a}ndez-Soto et al. (1999) and some
other previous studies (Gwyn \& Hartwick 1996; Lanzetta et al. 1996, 1998; Sawicki et al. 1997). 
However, we can identify such problematic galaxies by the additional
constraints explained in \S~\ref{sec:constraints} and reject them from
the sample to reduce systematic errors.

%\placefigure{fig:photoz}
%\placefigure{fig:chi2}

\begin{figure}
\epsscale{0.9}
\plotone{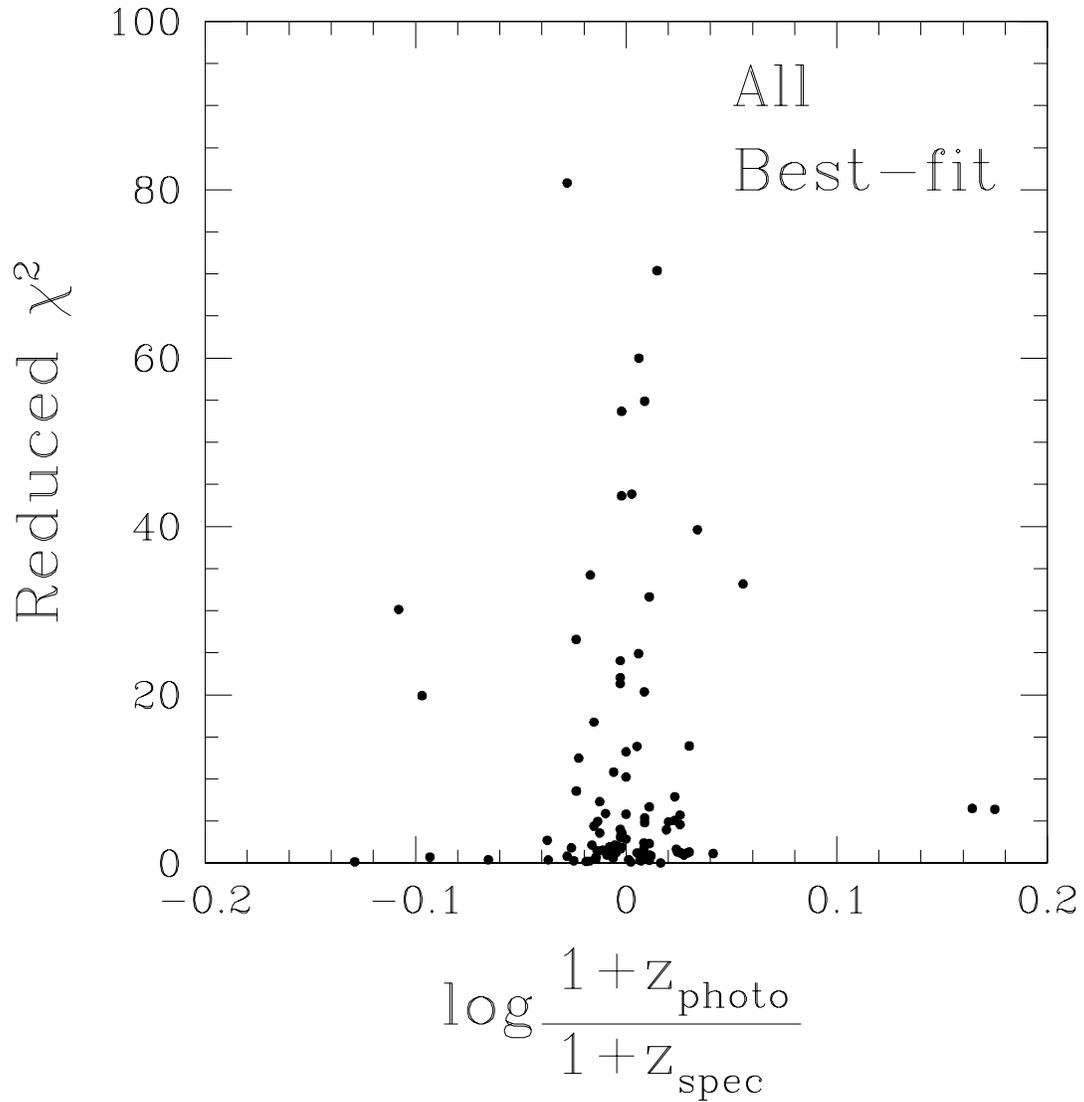}
\caption[fig6.ps]{Minimum reduced \chisq\  plotted against the error
 in redshift estimation for all galaxies in the spectroscopic
 sample.  The horizontal axis indicates the differences between 
\photoz\ and \specz\ in the logarithmic scale.}
\label{fig:chi2}
\end{figure}

We examine the minimum reduced \chisq\ value, which is an statistical 
indicator of the mean residual over {\it all} the observed bands.  
Figure~\ref{fig:chi2} shows the reduced \chisq\ values against the
differences between \photoz\ and \specz .  
Figure~\ref{fig:chi2} demonstrates that it is difficult to reject the
galaxies with bad \photoz\ based on the minimum reduced \chisq.
However, we find that the \chisq\ value {\it in the B band}, 
$\chi^2_{\rm B}$, is a good indicator for removing bad \photoz\
estimates.  We exclude 17 galaxies with $\chi^2_{\rm B} > 40$ from the sample.
The breakdown number of these galaxies for different redshift bins is
shown in Table~\ref{table:breakdown}. 

It is probably inappropriate to pursue 100\% success in \photoz,
especially for a sample of very distant galaxies.  It is inevitable that
some fraction of the sample fails to give correct \photoz.  The
important point is how to identify such false \photoz\ and reject them
from the sample so that the final sample has the least contamination
even with less completeness.  

\begin{figure}
\epsscale{0.8}
\plotone{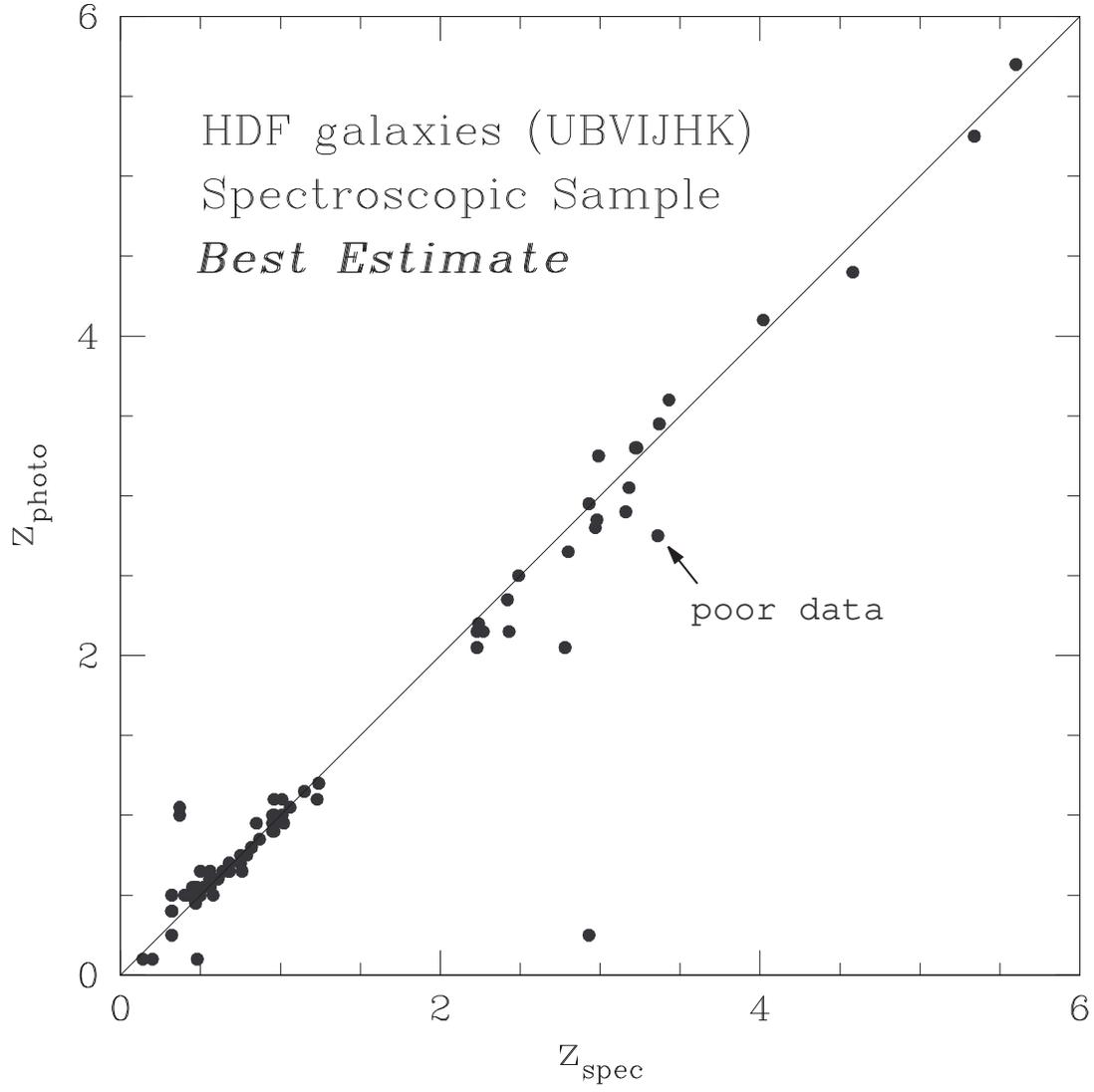}
\caption[fig7.ps]{Our adopted \photoz\ versus \specz\ for the spectroscopic
 sample.  The galaxy pointed by the arrow has a very poor photometric accuracy.}
\label{fig:photoz_best}
\end{figure}

We now have photometric redshifts for 91 out of 108 galaxies in the
spectroscopic sample, which are shown in Figure~\ref{fig:photoz_best}.
In Figure~\ref{fig:photoz_best}, the galaxy pointed by an arrow has
very poor S/N.  There are two galaxies with large deviations at
$z_{\rm spec}\simeq 0.5$.  Hogg et al. (1998) noted that the 
spectroscopic redshifts of these galaxies are determined by just one emission line, suggesting that the line identification is possibly wrong.
Table~\ref{table:sigma} presents the accuracy of our \photoz\ estimates
for the 91 galaxies.  We present in Table~\ref{table:sigma} the standard
deviations for three cases, total 91 galaxies, 90 galaxies without the
$z_{\rm spec}=2.93$ galaxy having a catastrophic error, and 86 galaxies
with $|z_{\rm spec}-z_{\rm photo}|\le 0.5$.
It is found that our method gives almost the same accuracy as previous
estimates for galaxies at $z<2$, and gives more accurate \photoz\ for
galaxies at $z>2$.

%\placefigure{fig:photoz_best}
%\placetable{table:breakdown}
%\placetable{table:sigma}

\begin{figure}
\epsscale{0.9}
\plotone{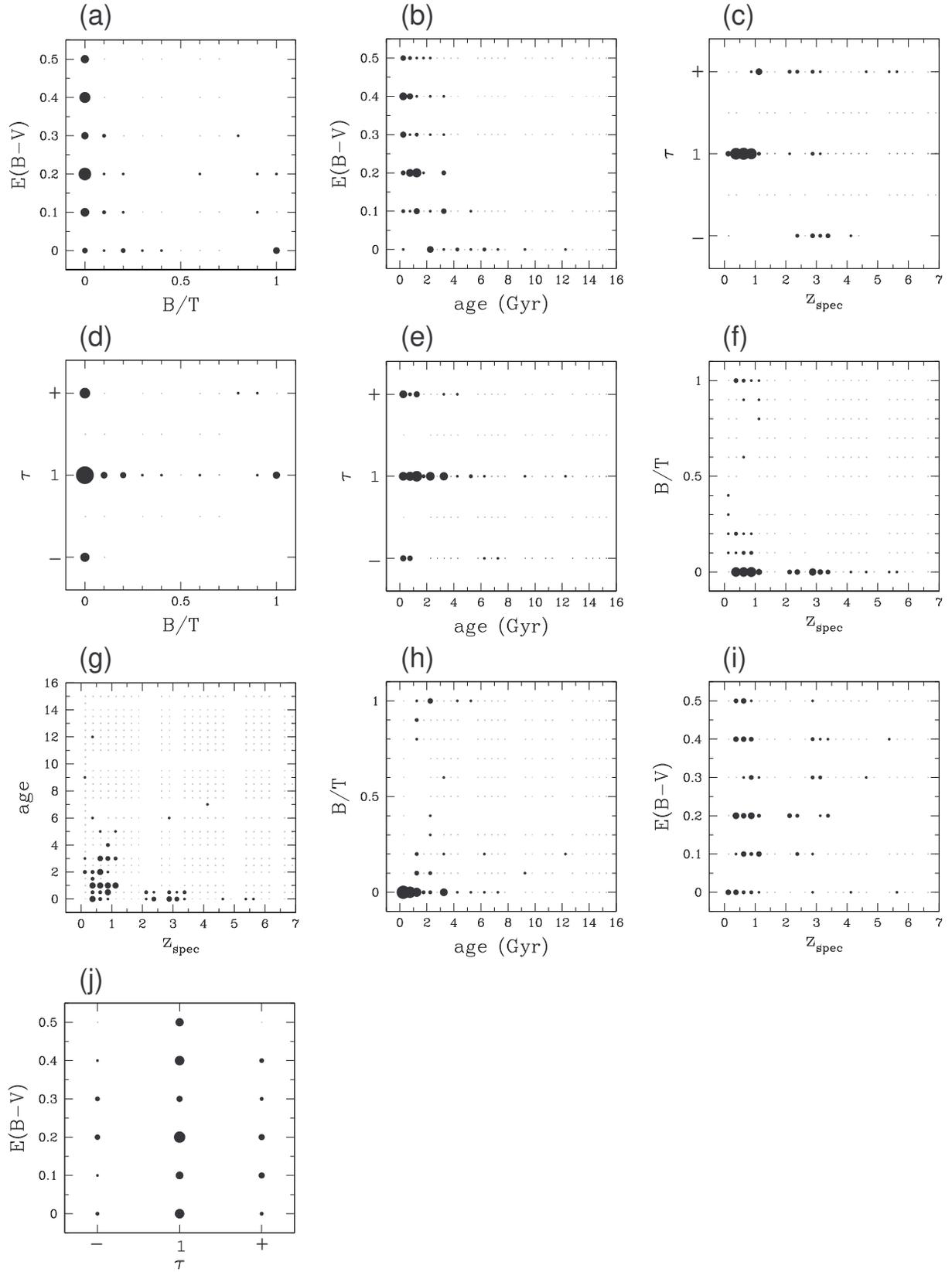}
\caption[fig8.ps]{The parameter distributions of the best-fit SEDs for the
 spectroscopic sample.  The area of the filled circles is proportional to 
the number of SEDs in the grid. For example, in panel (d) (B/T
 vs. $\tau$), the grid of (B/T,$\tau$)=(0.0, 1) includes 41 galaxies, 
the largest circle of all.  See the text for details.}
\label{fig:param_specz}
\end{figure}

In order to show that the \chisq\ minimizing method selects a {\it
reasonable} SED from the SED templates for each of the observed
galaxies, we present in Figure~\ref{fig:param_specz} `2-D scatter
diagrams' of parameters for the best-fit SEDs.  
%change
The area of the filled circle is proportional to the number of SEDs on
the grid.
In the panels, $\tau =1$ corresponds to the Madau value of $\tau_{\it
eff}$ and $\tau =-$ and $+$ represent $\tau_{\it eff}^-=0.5\tau_{\it
eff}$ and $\tau_{\it eff}^+=1.5\tau_{\it eff}$, respectively.
In panel (c), all galaxies at $z<1$ are fitted with $\tau =1$, but this is an artifact. For galaxies at $z<1$, the intergalactic Lyman absorption does
not enter any of the seven passbands, and $\tau =1$ is assigned to such galaxies.

Figure~\ref{fig:param_specz} demonstrates that the distributions of
parameters seem to be broadly consistent with the properties of actual 
galaxies.  For example, panel (a), the diagram of `B/T vs. $E(B-V)$', shows
that the population with large $E(B-V)$ is dominated by disk-dominated 
galaxies.  Also in panel (g), the `\specz\ vs. age' diagram, no old
galaxies are seen at high redshifts.  We conclude that
Figure~\ref{fig:param_specz} gives indirect support to the validity of
our method.

%\placefigure{fig:param_specz}

\begin{figure}
\epsscale{0.8}
\plotone{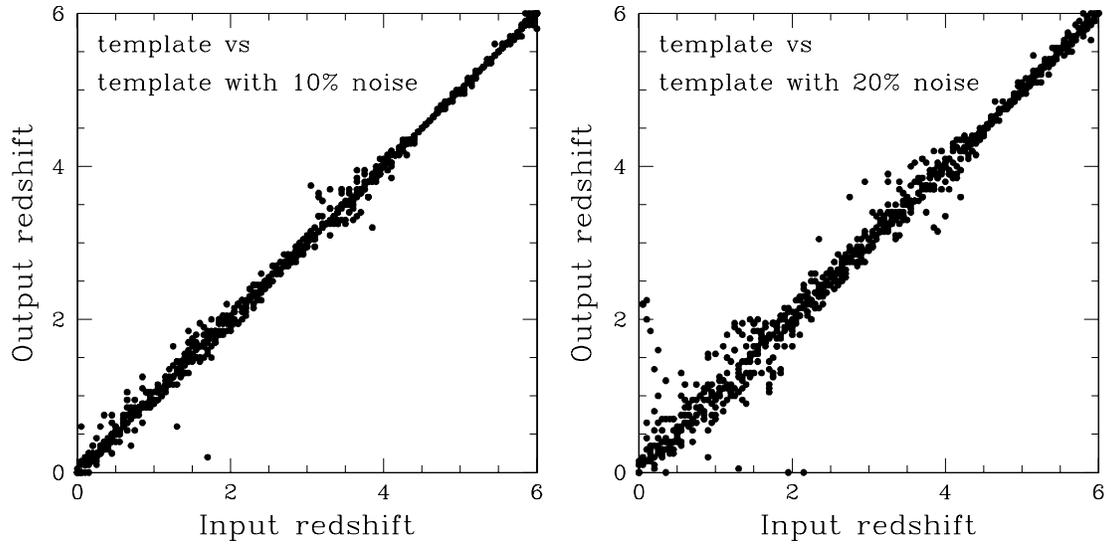}
\caption[fig9.ps]{Output \photoz\ versus input \specz\ of the simulation. 
The left and right panels are for the 10\% and 20\% input noises in flux, respectively.}
\label{fig:noise}
\end{figure}

%change
Finally, we examine the effects of different S/N on the accuracy of
\photoz\ using simulations.
We prepare 1,077 input galaxies by a random selection from total 
541,926 grids of our SED template, which consist of 3,366 SEDs and 161 
redshifts.  These input galaxies cover a wide range of spectral types.  
We add to them Gaussian noise so that photometric errors in {\it
UBVIJHK} bandpasses be $\Deltait =10$\%, which is roughly the same as
those of the HDF galaxies.  We obtain \photoz\ of these input galaxies 
in exactly the same way as for the HDF galaxies.  
The results are shown in Figure~\ref{fig:noise} (the case of $\Deltait
=20$\% is also shown for comparison) and the values of $\sigma_z$ are
summarized in Table~\ref{table:sigma_noise}.
The best-fit SED for a input galaxy is not always the
same as input SED itself because we add Gaussian noise to the magnitudes 
of input galaxies.  However, if we do not add noise, exactly the same 
SED as input is answered.

%\placefigure{fig:noise}
%\placetable{table:sigma_noise}

%%%%%%%%%%%%%%%%%%%%%%%%%%%%%%%%%%%%%%%%%%%%%%%%%%%%%%%%%%%%%%%%%%%%%%%%%
%                            Section 4                                  %
%%%%%%%%%%%%%%%%%%%%%%%%%%%%%%%%%%%%%%%%%%%%%%%%%%%%%%%%%%%%%%%%%%%%%%%%%

\section{COMPARISON WITH PREVIOUS WORK}\label{sec:comp}
%change
Hogg et al. (1998) reported a comparison among \photoz\ 
by various authors for the same low-{\it z} ($z<1.4$) galaxies in the
HDF.  %
We summarize their results (Table~3 of Hogg et al. 1998) again in
Table~\ref{table:hogg}, where we also give our result for exactly the same sample for reference.  %
It is found that the accuracy of our \photoz\ estimate for these
low-$z$ galaxies is comparable to those by the other authors.
However, it should be noted that the authors given in Hogg et al. (1998)
estimated their \photoz\ as a completely blind test, while our estimate
is not blind.  %
We have optimized some of the parameters in our method on the basis of
the photometry and spectroscopy catalog by Fern\'{a}ndez-Soto et al. (1999) which already includes most of the galaxies used for Hogg et al.'s blind test. 
In that point, our result can not be directly compared with those by the
other authors.  
The optimization in our method is, however, done using $z>2$ galaxies
only so that the accuracy of \photoz\ for these high-$z$ galaxies be 
better.  Accordingly, the accuracy of \photoz\ for such low-$z$
galaxies is not affected significantly by the optimization.

We compare our \photoz\ of high-$z$ galaxies in the HDF with those by 
Fern\'{a}ndez-Soto et al. (1999), which are based on 
the same sample and the same photometry as we use.  
Therefore, this is a direct comparison of the methods. 
Fern\'{a}ndez-Soto et al. had two galaxies with catastrophic errors, 
while we have one. 
The galaxy for which both of the authors give a catastrophic error is
located at $z_{\rm spec}=2.93$.  This galaxy is fitted by a young dusty 
spectrum of $0.5 {\rm Gyr}$ and $E(B-V)=0.5$, and cannot be rejected by
our \chisq\ threshold.  The other galaxy which has a catastrophic error
in Fern\'{a}ndez-Soto's study is rejected by the \chisq\ threshold in our
work, though it is also fitted by a dusty young disk SED. 
Table~\ref{table:soto} summarizes the \photoz\ accuracy for galaxies 
of $z>2$, with catastrophic errors excluded.
Our work gives a much improved accuracy for high-$z$ galaxies.
%change
Wang, Turner, \& Bahcall (1999) achieved $\sigma_z=0.08$ and $0.30$ for
$z<2$ and $2<z<4$ galaxies, respectively, which are comparable to our 
results.  They have adopted a training-set method which uses colors.

\begin{figure}
\epsscale{0.8}
\plotone{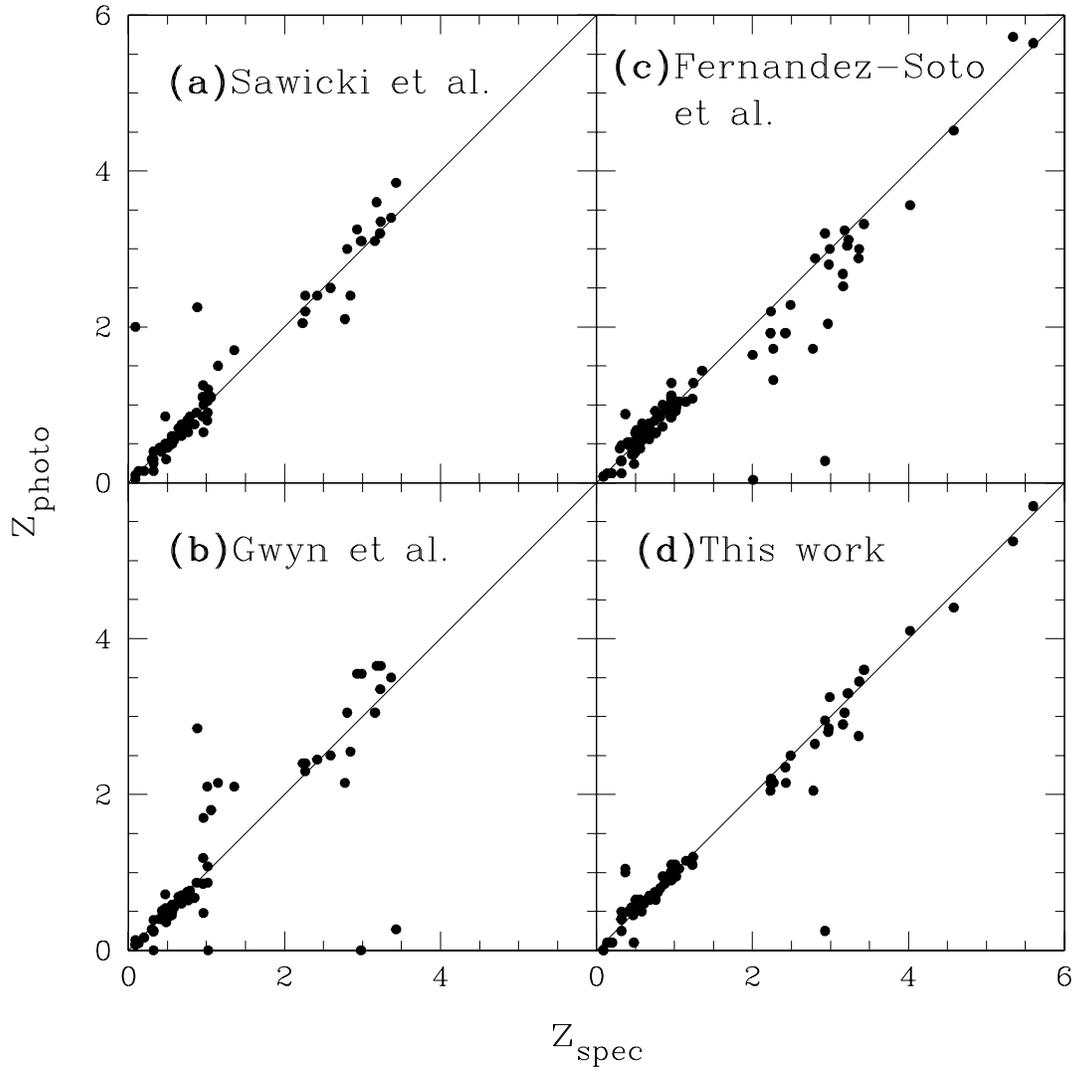}
\caption[fig10.ps]{Comparison of \photoz\ versus \specz\ among various authors. 
Results by Sawicki et al. and Gwyn et al. are based on optical 
4-band photometry for 74 galaxies with spectroscopic redshifts.
Results by Fern\'{a}ndez-Soto et al. and this work are based on 7-band 
photometry for 108 galaxies.}
\label{fig:comp}
\end{figure}

Figure~\ref{fig:comp} compares \photoz\ among authors (data of Gwyn \&
Hartwick (1996) and Sawicki et al. (1997) are taken from Sawicki [http://www.astro.utoronto.ca/\~{}sawicki/]). The results by
Fern\'{a}ndez-Soto et al. (1999) and ours have been newly added. 
The results by Sawicki et al. and Gwyn \& Hartwick are based on the 
optical 4-band photometry for 74 galaxies with spectroscopic redshifts,
while Fern\'{a}ndez-Soto et al.'s and our work are based on the 7-band photometry for 108 galaxies. 

Fern\'{a}ndez-Soto et al. and we obtain more accurate \photoz\ 
over all the redshift range than Gwyn et al. and Sawicki et al., presumably because 7-band photometry has longer baseline in 
wavelength and hence is more robust for estimating \photoz\ than 4-band photometry.
The optical 4-band photometry is worse at $z=1-2$ because in that
redshift range the 4000\AA\ break of galactic spectra drops out of the optical bands and no significant feature is available. 
The photometric redshifts by Gwyn \& Hartwick seem to be worse than 
Sawicki et al.'s.  This is probably because Sawicki et al. included
Lyman absorption into their template SEDs while Gwyn et al. did not.

%\placefigure{fig:comp}
%\placetable{table:hogg}
%\placetable{table:soto}

\begin{figure}
\epsscale{0.8}
\plotone{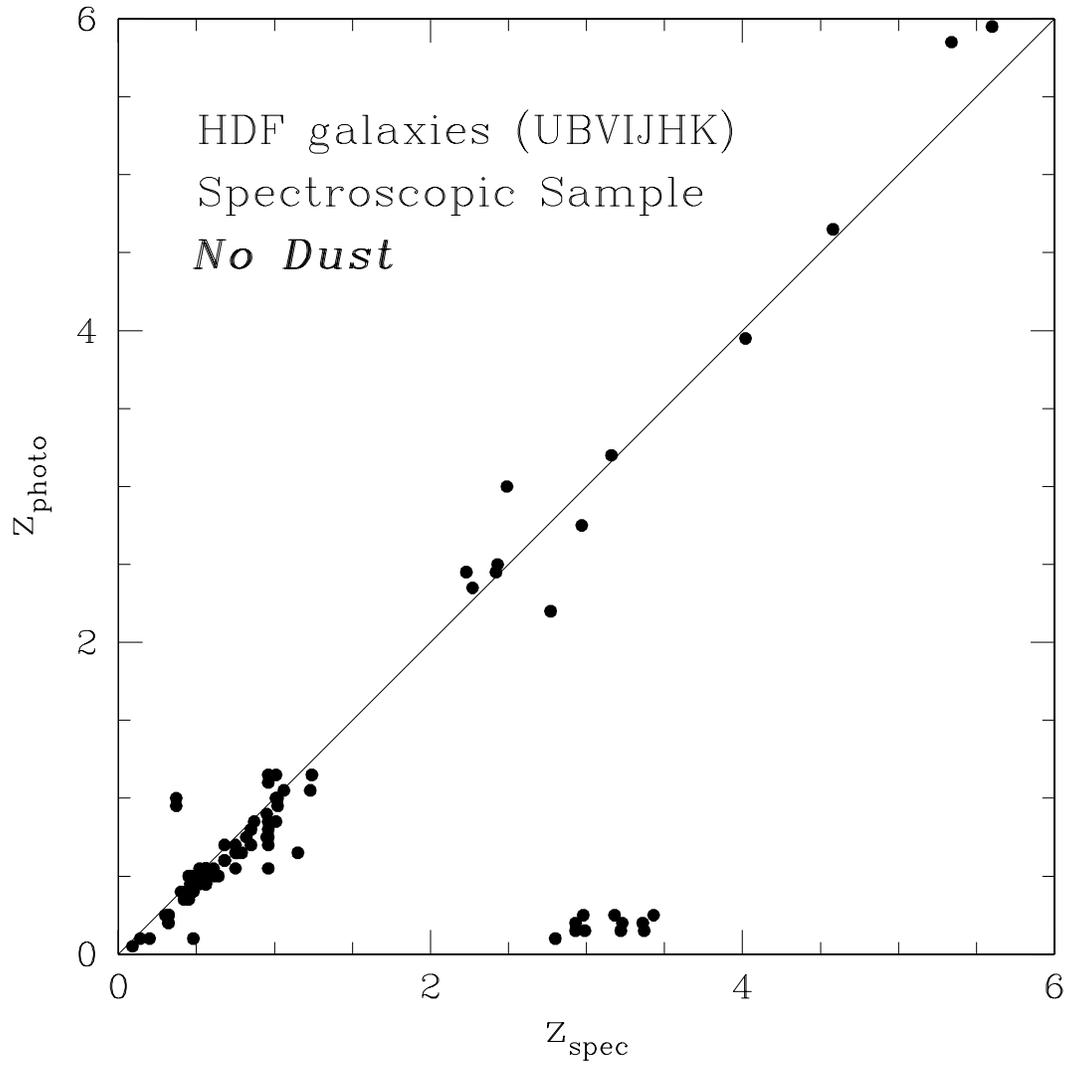}
\caption[fig11.ps]{\photoz\ based on the template SEDs without internal absorption. 
}
\label{fig:nodust}
\end{figure}

Now we examine the absorption effects on \photoz\ estimation.
In Gwyn et al.'s estimation, there are two galaxies with catastrophic
errors at $z_{\rm spec}\simeq 3$.
Sawicki et al. (1997) suggested that these catastrophic errors can be
improved by including the reddening effects into the template SEDs and
thereby reducing `{\it aliasing}', the misidentification between a 
4000\AA\ break and other breaks in shorter wavelengths.
We try our method using the template SEDs without internal
absorption.  The results are shown in Figure~\ref{fig:nodust}.
Figure~\ref{fig:nodust} indicates that the absence of internal absorption
in template SEDs increases catastrophic errors at $z_{\rm
spec}\simeq 3$. This is probably because `no dust SEDs' have steeper 4000\AA\ breaks than real observed galaxies, and therefore it is difficult to 
distinguish Lyman breaks (or some other breaks) from 4000\AA\ breaks
with no dust SEDs.
We have adopted dusty SEDs in this study.  The values of $E(B-V)$ are
taken in the range of 0.0 (no dust) to 0.5.
These values are consistent with those for Lyman-break galaxies at
$z=2-3$, $E(B-V)\simeq 0.3$, obtained by Sawicki \& Yee (1998). 
Calzetti \& Heckman (1999) also suggested $E(B-V)<0.5$ for galaxies at
$z>2-3$ by calculating the cosmic star formation rate and $E(B-V)$ 
values of galaxies iteratively.

\begin{figure}
\epsscale{0.8}
\plotone{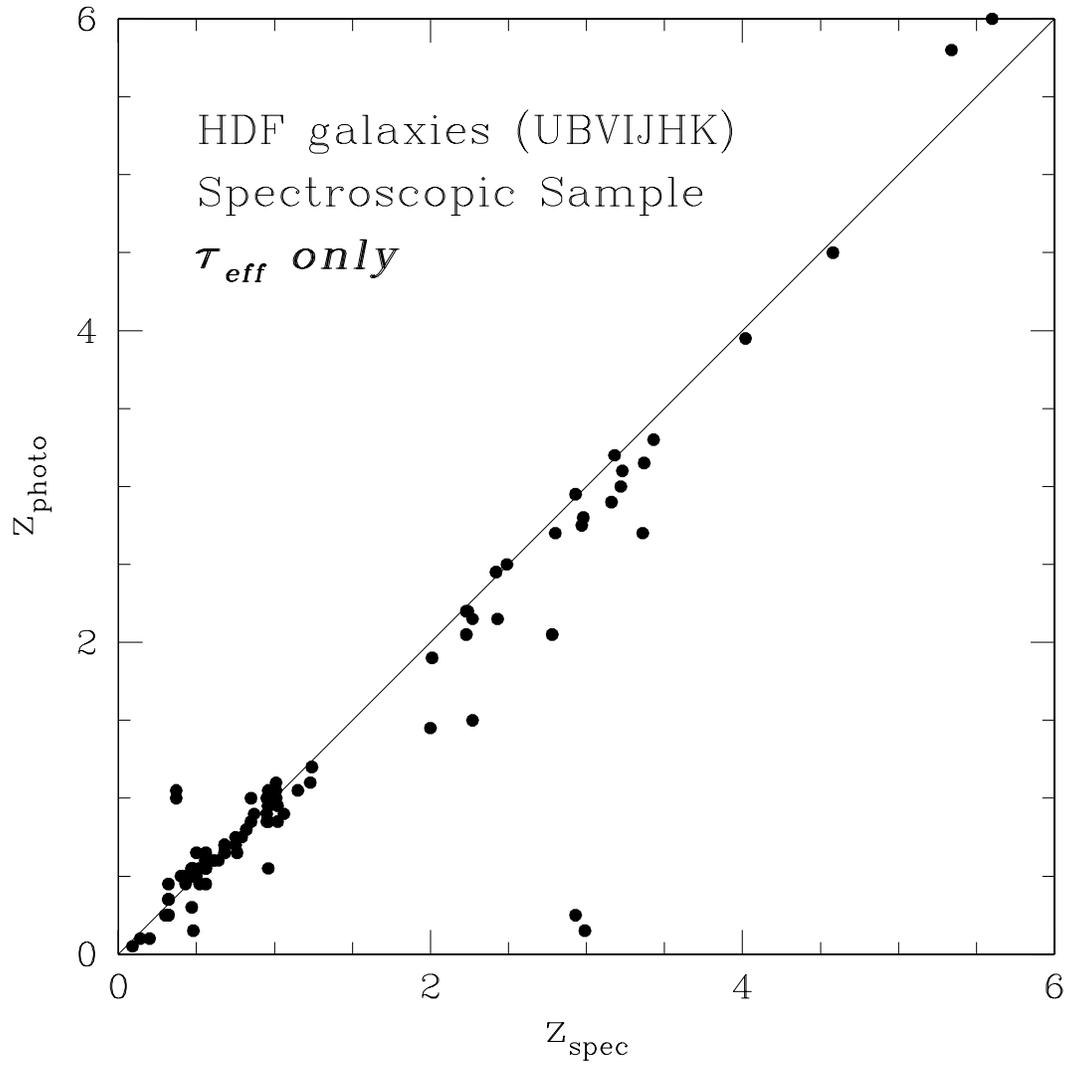}
\caption[fig12.ps]{\photoz\ based on the template SEDs using just one value,
 $\tau_{\it eff}$, for the opacity of intergalactic Lyman absorption.}
\label{fig:tau1}
\end{figure}

In Figure~\ref{fig:comp}(a)-(c), 
we see that galaxies at $z_{\rm spec}=2-4$, especially for the
seven-band photometry, tend to be generally estimated at lower redshifts
than \specz\ by the three previous studies.  
We are able to remove this systematic error by including three
values of opacity due to the intergalactic Lyman absorption around the
median opacity derived from Madau (1995):  We include $\tau_{\it eff}$, the exact value by Madau, $\tau_{\it eff}^{-}(\equiv 0.5\tau_{\it eff})$, and $\tau_{\it eff}^{+}(\equiv 1.5\tau_{\it eff})$. 
The systematic errors at $z=2-4$ emerge if we remove $\tau_{\it
eff}^{+}$ and $\tau_{\it eff}^{-}$ and use only $\tau_{\it eff}$ (Figure~\ref{fig:tau1}).
It is clear that including a variation of the opacity is essential.
We also examine the template SEDs with 3$\tau_{\it eff}$, 4$\tau_{\it
eff}$, 0.3$\tau_{\it eff}$, and 0.25$\tau_{\it eff}$, and find that the
best combination is ($\tau_{\it eff}$,$\tau_{\it eff}^+$,$\tau_{\it
eff}^-$).  We should point out that the ambiguities of the stellar
synthesis model used for the template SEDs, especially those in the UV
flux, may be coupled with the ambiguity of the internal absorption and
the statistical fluctuation of the intergalactic Lyman absorption.  
If we look at Figure~\ref{fig:param_specz}(c), best-fit $\tau$ values 
seem to change with redshift: $\tau_{\it eff}^{+}$ is more
favorable for \specz$=2-4$ and $\tau_{\it eff}^{-}$ is more favorable
for \specz$>4$.
However, we do not insist that this trend is real considering the fact
that we do not have much knowledge about the UV flux of galaxies.
Nonetheless, we can obtain more accurate \photoz\ at $2<z<4$ by 
carefully treating the intergalactic Lyman absorption.

In summary, we find that the internal absorption is effective in
reducing the catastrophic errors, while the intergalactic absorption is 
effective in reducing the systematic error at $z=2-4$. 
 
%\placefigure{fig:nodust}
%\placefigure{fig:tau1}

%%%%%%%%%%%%%%%%%%%%%%%%%%%%%%%%%%%%%%%%%%%%%%%%%%%%%%%%%%%%%%%%%%%%%%%%%
%                            Section 5                                  %
%%%%%%%%%%%%%%%%%%%%%%%%%%%%%%%%%%%%%%%%%%%%%%%%%%%%%%%%%%%%%%%%%%%%%%%%%

\section{PROPERTIES OF THE HDF GALAXIES BASED ON THE IMPROVED
PHOTOMETRIC REDSHIFTS}\label{sec:property}

\subsection{How Many Template SEDs?}\label{sec:how_many}
We have used 187 template SEDs covering a wide range of spectral types. 
Generally speaking, accurate \photoz , even for high-$z$ galaxies, 
can be obtained using either observed spectra (CWW; Kinney et
al. 1993) or simulated spectra (KA97; BC96), though absorption 
effects (internal and intergalactic absorption) must be carefully 
treated.
If the purpose of the analysis is just to obtain \photoz\ (e.g.,
Fern\'{a}ndez-Soto et al. 1999), a few template SEDs would be sufficient.
However, if we want to determine accurately the SED shape of a target
galaxy in order to derive useful information on the SED, much more template SEDs are necessary as we will see below. 

\begin{figure}
\epsscale{0.9}
\plotone{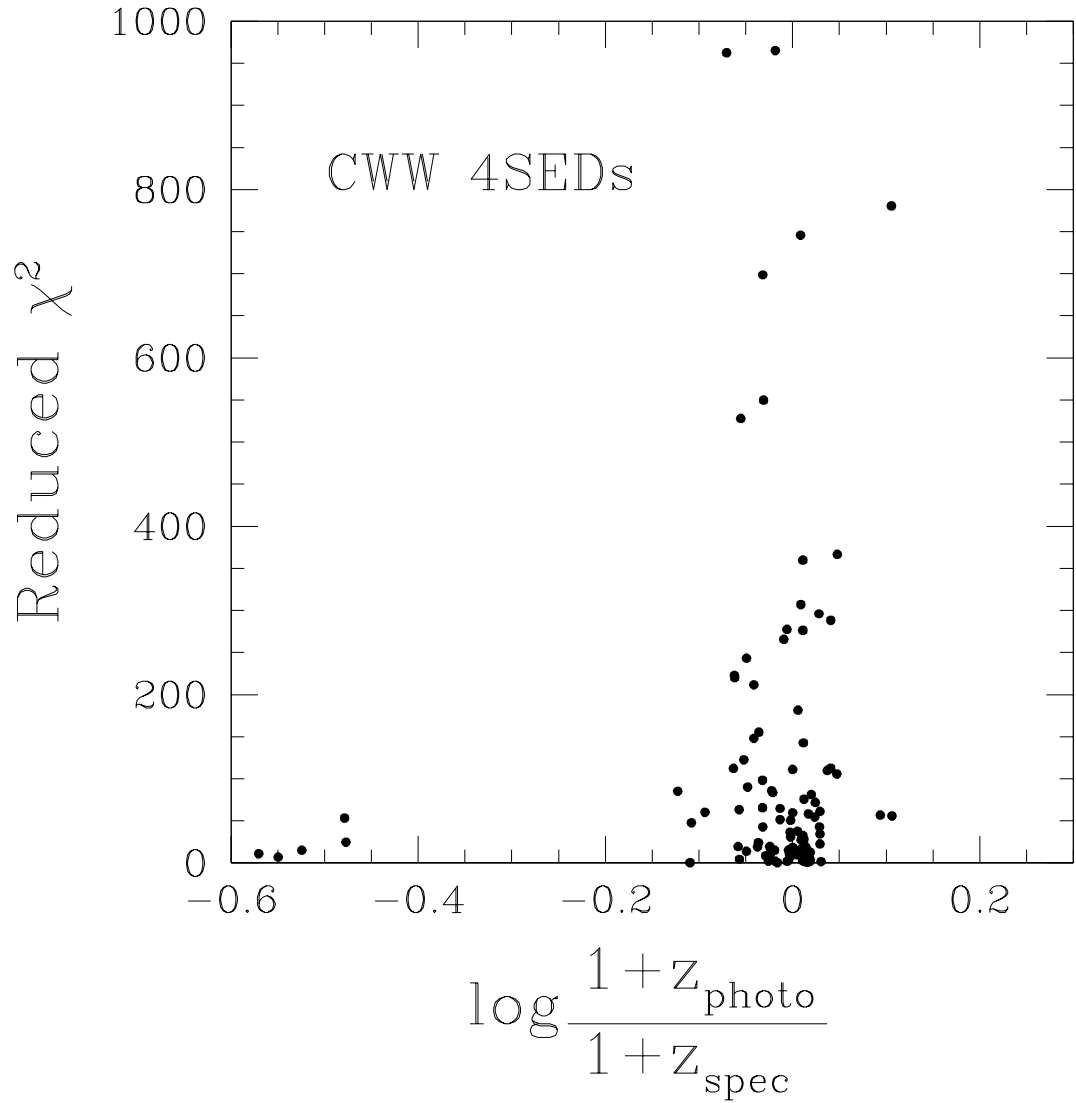}
\caption[fig14.ps]{Minimum reduced \chisq\ obtained with the CWW's four SEDs as a 
 template set. Compare with Figure~\ref{fig:chi2}.}
\label{fig:chi2_CWW4}
\end{figure}

Figure~\ref{fig:photoz_CWW4} shows \photoz\ of the HDF galaxies using 
four SEDs from CWW as template SEDs, which is almost the same procedure as Fern\'{a}ndez-Soto et al. adopted (Note that we reproduce the same 
trend in Figure~\ref{fig:photoz_CWW4} as seen in Figure~\ref{fig:comp}(c)). 
The accuracy of \photoz\ based on the four SEDs is worse than that of
our best estimate (Figure~\ref{fig:comp}(c)), but the difference
is within a factor of about 2.
However, there is a large difference in the minimum reduced \chisq. 
Our minimum reduced \chisq\ values are plotted in Figure~\ref{fig:chi2}.
Figure~ \ref{fig:chi2_CWW4} is the same plot as Figure~\ref{fig:chi2}
but for the four-SED case. The minimum reduced \chisq\ values obtained
with the four-SED templates are much larger than those with 3,366
($187\times 6 E(B-V) \times 3\tau $) templates. The distribution of
reduced \chisq\ values can be a measure to determine how many template
SEDs are necessary and sufficient for a given purpose.

%\placefigure{fig:photoz_CWW4}
%\placefigure{fig:chi2_CWW4}

\subsection{Properties of the HDF Galaxies}
\begin{figure}
\epsscale{0.8}
\plotone{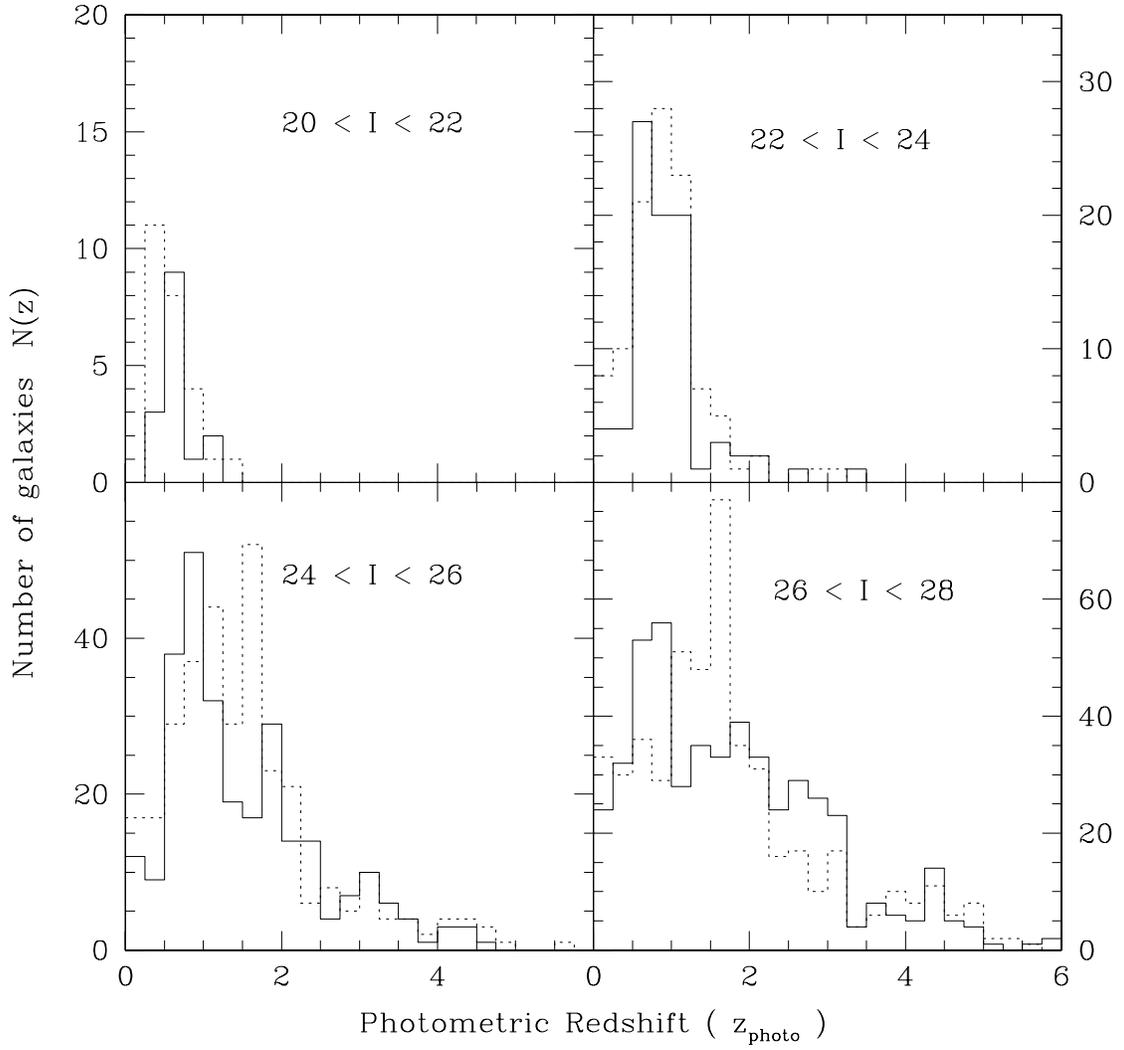}
\caption[fig15.ps]{$N(z)$ of the photometric sample. The solid and dotted lines
 indicates our results and Fern\'{a}ndez-Soto et al.'s.}
\label{fig:compare_nz} 
\end{figure}

We obtain photometric redshifts of 925 galaxies out of the 946 galaxies
in the good S/N region (See \S~\ref{sec:data}).  We call these 925
galaxies the photometric sample.
The redshift distribution $N(z)$ of the photometric sample is shown in
Figure~\ref{fig:compare_nz}.  For comparison, the results by
Fern\'{a}ndez-Soto et al. are superposed. 

Sawicki et al. (1997) argued that $N(z)$ could be an indicator of the
error in photometric redshift technique.  
They claimed that {\it aliasing} among spectral breaks (4000\AA, Balmer,
2800\AA, and 2635\AA) easily leads to an unrealistic $N(z)$, which has
typically two prominent peaks at $z\simeq 0-1$ and $z\simeq 2$. 
The peaks are likely caused by a template set which does not include
the internal or the intergalactic absorption. 
Our $N(z)$ does not have such two prominent peaks. Instead, our $N(z)$ 
shows a single peak at $z\simeq 0.5-1$ and a moderate decrease at $z>1$ 
at all magnitude bins.

Fern\'{a}ndez-Soto et al. found that galaxies at $z=2-4$ have
systematically lower photometric redshfits (Figure~\ref{fig:comp}(c)).
We find in this paper that this is attributed to a systematic error 
which is sensitive to the intergalactic absorption.
This `small aliasing' due to the systematic error would make an 
unlialistic peak at $z\simeq 2$ in $N(z)$, especially for faint
galaxies. 
For $26<I<28$, the $N(z)$ of Fern\'{a}ndez-Soto et al. has a prominent
peak at $z\simeq 2$. This is probably due to the small aliasing of
galaxies at $z\ge 2$, because the number of Fern\'{a}ndez-Soto et al.'s
galaxies with $z>2$ is smaller than that of our galaxies.

%\placefigure{fig:compare_nz}

\begin{figure}
\epsscale{0.8}
\plotone{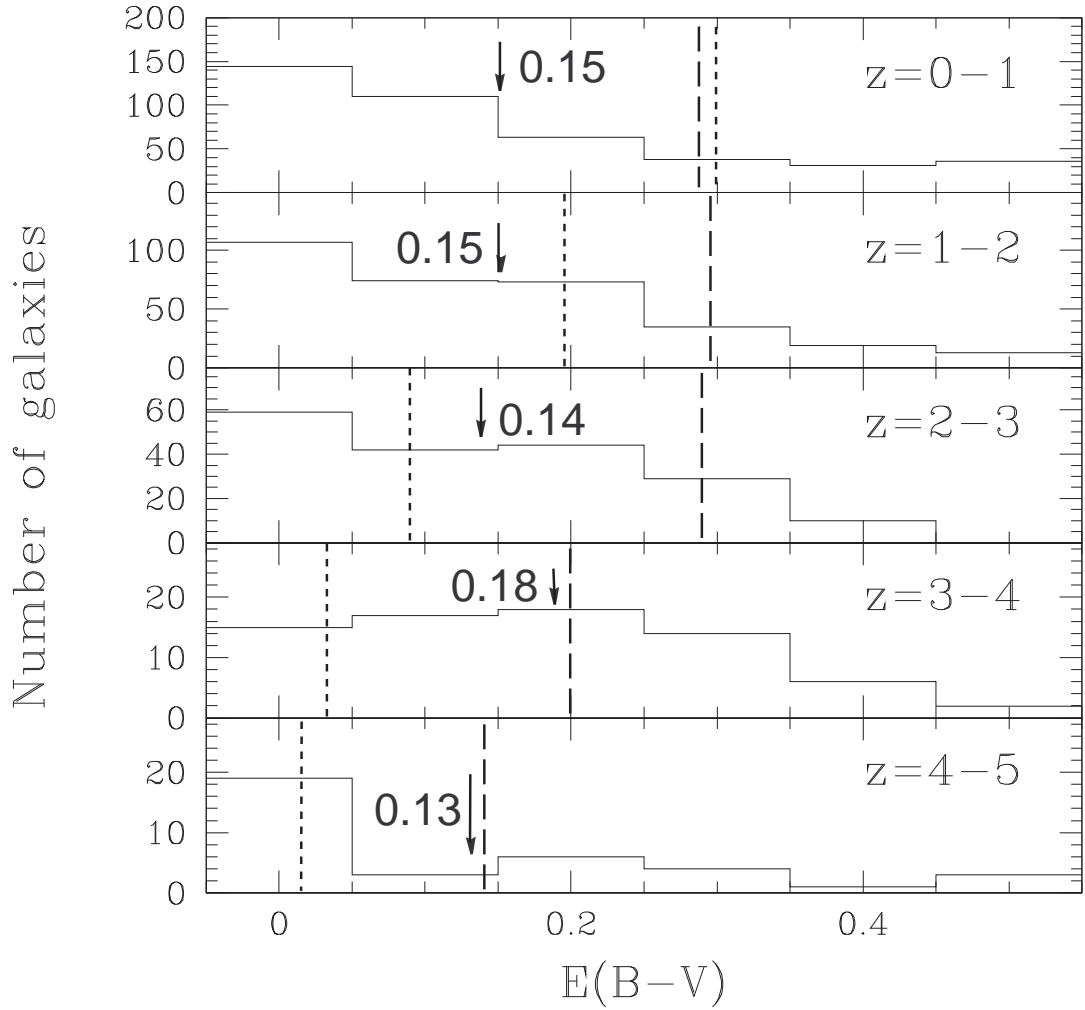}
\caption[fig16.ps]{Histograms of $E(B-V)$ for five redshift bins. 
Arrows indicate the mean of $E(B-V)$ in each redshift bin. Vertical
lines in each bin denote the predictions by Calzetti \& Heckman (1999). 
The dashed lines are for model `A' and the dotted lines are for 
model `B' (See text for details).}
\label{fig:dust_z}
\end{figure}

Next, we examine $E(B-V)$ of galaxies. Figure~\ref{fig:dust_z}  shows
the histograms of $E(B-V)$ of the best-fit SEDs in the five redshift
ranges. The arrow in each redshift range indicates the mean value.
We find that the $E(B-V)$ value of galaxies remains almost
constant ($E(B-V)=0.13-0.18$) for $z=0-6$, though it has a small peak at  
$z=3-4$ bin.  These values are roughly consistent with those of 
Lyman-break galaxies at $z=2.5-3.5$ derived by Sawicki \& Yee (1998;
$E(B-V)\simeq 0.3$), because our photometric sample probably 
includes not only star-forming Lyman-break galaxies but also passively
evolving elliptical galaxies. 
Calzetti \& Heckman (1999) iteratively calculated the cosmic star formation
rate (SFR) and the mean $E(B-V)$ of galaxies using the Galactic conversion
factor between the gas column density $N(H)$ and $E(B-V)$, i.e., 
$E(B-V)=N(H)/5.9\times 10^{21}$ (Bohlin, Savage, \& Drake 1978).
The vertical lines plotted in Figure~\ref{fig:dust_z} are
predictions of $E(B-V)$ based on their two typical models.  
The model `A' corresponds to the cosmic SFR which remains almost
constant from $z=1.5$ to $5$.  The model `B' corresponds to the cosmic
SFR which reaches its peak at $z=1-2$ and monotonically decreases at
$z=2-5$.  Our results are found to be more consistent with the model `A'.

%\placefigure{fig:dust_z}

%%%%%%%%%%%%%%%%%%%%%%%%%%%%%%%%%%%%%%%%%%%%%%%%%%%%%%%%%%%%%%%%%%%%%%%%%
%                            Section 5                                  %
%%%%%%%%%%%%%%%%%%%%%%%%%%%%%%%%%%%%%%%%%%%%%%%%%%%%%%%%%%%%%%%%%%%%%%%%%

\section{SUMMARY AND CONCLUSIONS}
We improve photometric redshifts of 1048 galaxies in the HDF. 
A standard \chisq\ minimizing method is adopted for obtaining \photoz.
We use 187 template SEDs representing a wide variety of 
morphology and age of observed galaxies based on a population 
synthesis model by Kodama \& Arimoto (1997) with two new recipes.
First, the amount of the internal absorption is changed as a free
parameter in the range of $E(B-V)=0.0$ to $0.5$ with an interval of $0.1$.  
Second, the absorption due to intergalactic HI clouds 
is also changed by a factor of 0.5, 1.0, and 1.5 around the
opacity given by Madau (1995). The total number of template SEDs
is thus $187\times 6\times 3=3,366$, except for the redshift grid.
The introduction of the internal absorption is found to be effective in
reducing the catastrophic errors in \photoz\ while the absorption due to
the intergalactic HI clouds helps to improve the systematic error in
\photoz\ at $z=2-4$.
The dispersion $\sigma_z$ of our photometric redshifts with respect 
to spectroscopic redshifts is $\sigma_z=0.08$ and $0.24$ for $z<2$ and
$z>2$, respectively, which are smaller than the corresponding values
($\sigma_z=0.09$ and $0.45$) by Fern\'{a}ndez-Soto et al. (1999). 
Significant improvements are obtained, especially at high-$z$.  
A comparison with previous work and the properties of the best-fit SEDs
for the spectroscopic sample verify the validity of our photometric redshifts.

The redshift distribution of all the 925 galaxies in the photometric sample 
shows a peak at $z\simeq 1$ and a moderate decrease at $z>1$
without any other peak. The $E(B-V)$ value of galaxies
remains almost constant for $z=0-6$ with a possible weak peak at $z=3-4$,
which is consistent with the model of a constant cosmic SFR given by
Calzetti \& Heckman (1999).

%\bigskip
\acknowledgments

We would like to express our gratitude to T. Kodama \& N. Arimoto for
kindly allowing us to use their new population synthesis code and to 
anonymous referee for constructive comments.  %
H.F. wishes to thank the Japan Society for the Promotion of Science for a
financial support.  This work is supported in part by Grants-in-Aid
(07CE2002, 11640228, 10440062) from the Ministry of Education, Science,
Sports and Culture of Japan.

\clearpage

%%%%%%%%%%%%%%%%%%%%%%%%%%%%%%%%%%%%%%%%%%%%%%%%%%%%%%%%%%%%%%%%%%%%%%%%%
%                      Table and Figure                                 %
%%%%%%%%%%%%%%%%%%%%%%%%%%%%%%%%%%%%%%%%%%%%%%%%%%%%%%%%%%%%%%%%%%%%%%%%%

\begin{deluxetable}{lccccl}
\footnotesize
\tablecaption{Parameters of template SEDs adopted in this work
\label{table:sedparam}
}
\tablewidth{0pt}
\tablehead{
 \colhead{SED type} & \colhead{$x^{\rm IMF}$\tablenotemark{(a)}} & 
 \colhead{$\tau_{\rm SF}$ [Gyr]} & \colhead{$\tau_{\rm infall}$ [Gyr]} & 
 \colhead{$t_{\rm GW}$ [Gyr]\tablenotemark{(b)}} & \colhead{Age[Gyr]}
}
\startdata
{}   & {}   & {}  & {}  & {}   &  0.010, 0.013, 0.016, 0.020, 0.025,\nl
{}   & {}   & {}  & {}  & {}   &  0.032, 0.040, 0.050, 0.063, 0.790,\nl
pure disk&1.35&5.0&5.0&20.0    &  0.100, 0.126, 0.158, 0.200, 0.251,\nl
(37kinds)&{}& {}  & {}  & {}   &  0.316, 0.398, 0.501, 0.631, 0.794,\nl
{}   & {}   & {}  & {}  & {}   &  1.0, 1.259, 1.585, 2.0, 3.0, 4.0, \nl
{}   & {}   & {}  & {}  & {}   &  5.0, 6.0, 7.0, 8.0, 9.0, 10.0, 11.0,\nl
{}   & {}   & {}  & {}  & {}   &  12.0, 13.0, 14.0, 15.0\nl
\tableline
pure bulge   & {}   & {}  & {}  & {}    & 1.0, 2.0, 3.0, 4.0, 5.0,\nl
(15kinds)    & 1.10 & 0.1 & 0.1 & 0.353 & 6.0, 7.0, 8.0, 9.0, 10.0,\nl
{}           & {}   & {}  & {}  & {}    & 11.0, 12.0, 13.0, 14.0, 15.0\nl
\tableline
Bulge+Disk\tablenotemark{(c)} & {}   & {}  & {}  & {}    & 1.0, 2.0, 3.0, 4.0, 5.0,\nl
(9$\times$15kinds)  & {$\cdots$}  & {$\cdots$} & {$\cdots$} & {$\cdots$}   & 6.0, 7.0, 8.0, 9.0, 10.0,\nl
{}& {} & {} & {} & {}   & 11.0, 12.0, 13.0, 14.0, 15.0\nl
\enddata
\tablenotetext{(a)}{Power-law index of initial mass function.}
\tablenotetext{(b)}{$t_{\rm GW}=20$ means that the galactic wind does not
 blow until the present epoch.}
\tablenotetext{(c)}{B/T=0.1-0.9, where B and T are the bulge and the
 total luminosity.}
\end{deluxetable}

\begin{deluxetable}{cccc}
\footnotesize
\tablecaption{The photometry of the HDF\label{table:obs_hdf}}
\tablewidth{0pt}
\tablehead{
 \colhead{Band} & \colhead{Exposure time} &
 \colhead{PSF(FWHM)} & \colhead{limiting magnitude}\\
 \colhead{} & \colhead{[hours]} & \colhead{[arcsec]} & \colhead{[mag]}
}
\startdata
F300W({\it U}) & 42.7      & $\cdots$ & 26.98(AB) for $10\sigma$ limit \nl
F450W({\it B}) & 33.5      & $\cdots$ & 27.86(AB) for $10\sigma$ limit \nl
F606W({\it V}) & 30.3      &  0.12  & 28.21(AB) for $10\sigma$ limit \nl
F814W({\it I}) & 34.3      & $\cdots$ & 27.60(AB) for $10\sigma$ limit \nl
{\it J}        & 11.0      &  1.0    & 23.45(STD) for $5\sigma$ limit \nl
{\it H}        & 11.3      &  1.0    & 22.29(STD) for $5\sigma$ limit \nl
{\it K}        & 22.9      &  1.0    & 21.92(STD) for $5\sigma$ limit \nl
\enddata
\tablecomments{Upper 4 rows: HST-WFPC2 ({\it UBVI}) (Wiiliams et al. 1996) and lower 3 rows: KPNO-IRIM ({\it JHK}) (Dickinson et al. 1999).}
\end{deluxetable}

\begin{deluxetable}{cccc}
\footnotesize
\tablecaption{The breakdown number of galaxies in the spectroscopic sample
\label{table:breakdown}
}
\tablewidth{0pt}
\tablehead{
 \colhead{\specz} & \colhead{0.0 - 6.0} & \colhead{0.0 - 2.0} & 
 \colhead{2.0 - 6.0} 
}
\startdata
Total with \specz    &   108   &   79    &   29           \nl
Good[$\chi^2_{\rm B}<40$]   &91(84.3) &66(83.5) &25(86.2) \nl
Bad[$\chi^2_{\rm B} \ge 40$]&17(15.7) &13(16.5) &4(13.8)  \nl
\enddata
\tablecomments{The values in the brackets mean the ratio(\%) to the total number within each redshift bin.}
\end{deluxetable}

\begin{deluxetable}{cccc}
\footnotesize
\tablecaption{Summary of accuracy of our \photoz\ for the spectroscopic
 sample
\label{table:sigma}
}
\tablewidth{300pt}
\tablehead{
 \colhead{\specz} & \colhead{Number of galaxies} & \colhead{$\sigma_z$}
}
\startdata
{}           & [1]  91         & 0.329\nl
0.0 - 6.0    & [2]  90         & 0.171\nl
{}           & [3]  86         & 0.101\nl\tableline
{}           & [1]  66         & 0.139\nl
0.0 - 2.0    & [2]  66         & 0.139\nl 
{}           & [3]  64         & 0.081\nl\tableline
{}           & [1]  25         & 0.585\nl
2.0 - 6.0    & [2]  24         & 0.239\nl
{}           & [3]  22         & 0.145\nl
\enddata
\tablenotetext{}{[1] For all galaxies.}
\tablenotetext{}{[2] One galaxy with a catastrophic error is removed.}
\tablenotetext{}{[3] Galaxies of $|$\specz - \photoz $| > 0.5$ are removed.}
\end{deluxetable}

\begin{deluxetable}{cccc}
\footnotesize
\tablecaption{Summary of accuracy of our \photoz\ for the simulation with 10\% Gaussian noise
\label{table:sigma_noise}
}
\tablewidth{0pt}
\tablehead{
 \colhead{\specz} & \colhead{0.0 - 6.0} & \colhead{0.0 - 2.0} &
 \colhead{2.0 - 6.0}
}
\startdata
Number of input galaxies & 1077 (1071)   &  272 (269)    & 805 (802) \nl
$\sigma_z$  & 0.102 (0.081) & 0.157 (0.116) & 0.076 (0.066) \nl
\enddata
\tablecomments{Numbers in the parenthesis are for those with $|$\specz - \photoz $|<0.5$.}
\end{deluxetable}

\begin{deluxetable}{ccccc}
\footnotesize
\tablecaption{Comparison of accuracies for galaxies with
 $0<$\specz$<1.4$ according to Hogg et al. (1998)
\label{table:hogg} 
}
\tablewidth{0pt}
\tablehead{
 \colhead{Author} & \colhead{Number of galaxies} & \colhead{Band} & \multicolumn{2}{c}{Fraction\tablenotemark{1)}}\\
 \colhead{} & \colhead{} & \colhead{} & \colhead{$\Deltait z\le 0.1$[\%]} & \colhead{$\Deltait z\le 0.3$[\%]}
}
\startdata
Gwyn           & 23  & {\it UBVI}    & 65.2    &  91.3 \nl
Mobasher       & 20  & {\it UBVI}    & 35.0    &  80.0 \nl
Sawicki-4band  & 20  & {\it UBVI}    & 65.0    &  85.0 \nl
Sawicki-7band  & 20  & {\it UBVIJHK} & 70.0    &  95.0 \nl
Connolly       & 20  & {\it UBVIJ}   & 65.0    &  95.0 \nl
Fern\'{a}ndez-Soto & 19  & {\it UBVIJHK} & 73.7  &  94.7 \nl
\tableline
This Work\tablenotemark{2)}      & 19  & {\it UBVIJHK} & 78.9    &  94.7 \nl
\enddata
\tablenotetext{1)}{Fraction (\%) of galaxies which have $\Deltait z<0.1$ and $0.3$ is given for different authors.}
\tablenotetext{2)}{This work is not a blind test as explained in the
 test.  However, the same sample is used as in Fern\'{a}ndez-Soto's result.}
\end{deluxetable}

\begin{deluxetable}{ccc}
\footnotesize
\tablecaption{Comparison of $\sigma_z$ of high-$z$ ($z>2$) galaxies between our estimates and Fern\'{a}ndez-Soto et al.'s
\label{table:soto}
}
\tablewidth{0pt}
\tablehead{
 \colhead{} & \colhead{$\sigma_z$ (Fern\'{a}ndez-Soto)} & \colhead{$\sigma_z$ (This work)}
}
\startdata
Sample A (N=27) &    0.58   &   0.29  \nl
Sample B (N=26) &    0.45   &   0.30  \nl
Sample C (N=24) &    0.42   &   0.24  \nl
\enddata
\tablecomments{Numbers in the parenthesis are number of galaxies in
 each sample.  [A]: A galaxy which has a catastrophic error in our
 estimate is rejected.  [B]: Two galaxies which have a catastrophic
 error in Fern\'{a}ndez-Soto et al's estimate are rejected.  [C]: Five galaxies which have either a catastrophic error or a large \chisq\ value in our estimate are rejected.}
\end{deluxetable}

\clearpage

\begin{figure}
\epsscale{0.8}
\plotone{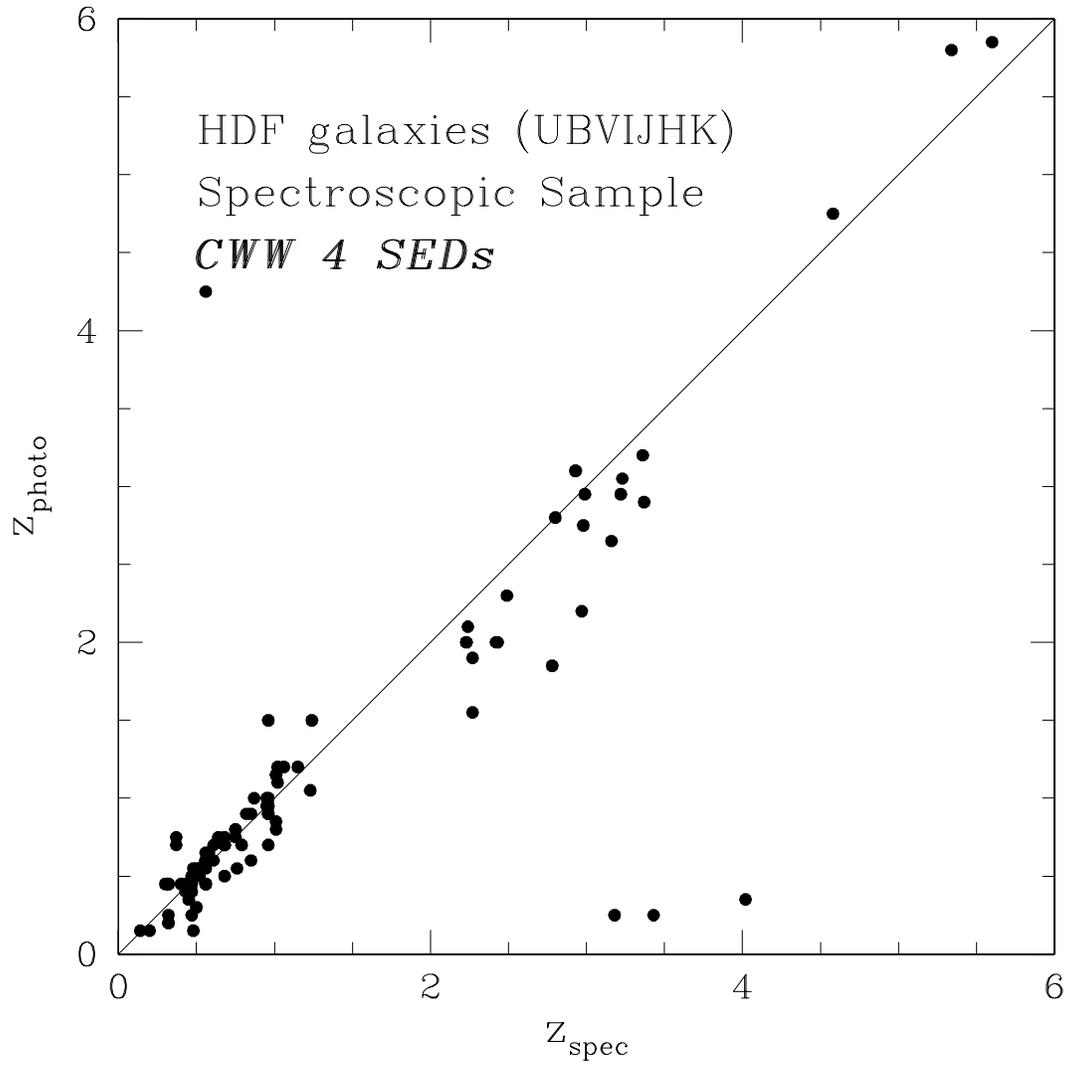}
\caption[fig13.ps]{\photoz\ obtained based on CWW's four SEDs as a template set.}
\label{fig:photoz_CWW4}
\end{figure}

\clearpage


\begin{thebibliography}{}
%\bibitem[Anders et al.,\ 1989]{and89}
%Anders, E., and Grevesse, M. 1989. Geochim. Cosmochim. Acta, 53, 197

\bibitem[Baum et al. 1963]{baum1963} 
Baum, W. 1963, IAU Sym. 15: Problems of Extragalactic Research, (New York; Macmillan), p. 390

\bibitem[Bianchi et al. 1996]{bianchi1996}
Bianchi, L., Clayton, G. C., Bohlin, R. C., Hutchings, J. B., \& Massey, P. 1996, \apj, 471, 203

\bibitem[Bouchet et al. 1985]{bouchet1985} 
Bouchet, P., Lequeux, J., Maurice, E., Pr\'{e}vot, L., \& Pr\'{e}vot-Burnichon, M. L. 1985, A\&A, 149, 330

\bibitem[Bohlin, Savage, \& Drake 1978]{bohlin1978} 
Bohlin, R. C., Savage, B. D., \& Drake, J. F. 1978, \apj, 224, 132

\bibitem[Bruzual \& Charlot 1993]{bc93}
Bruzual, A. G. \& Charlot, S. 1993, \apj, 405, 538

\bibitem[Bruzual \& Charlot]{bc96} 
Bruzual, A. G. \& Charlot, S. 1996, GISSEL96, \apj\ in preparation

\bibitem[Cardelli, Clayton, \& Mathis 1989]{cardelli1989} 
Cardelli, J. A., Clayton, G. C., \& Mathis, J. S. 1989, \apj, 345, 245

\bibitem[Carlberg et al. 1998]{carlberg1998} 
Carlberg R. G., Yee, H. K. C., Morris, S. L., Lin, H., Sawicki, M., Wirth, G., Patton, D., Shepherd, C. W., Ellingson, E., Schade, D., Pritchet, C. J., \& Hartwick, F. D. A. 1998, astro-ph/9805131

\bibitem[Calzetti 1997a]{calzetti1997a} 
Calzetti, D. 1997, \aj, 113, 162

\bibitem[Calzetti 1997b]{calzetti1997b} 
Calzetti, D. 1997, uulh.conf., 403

\bibitem[Calzetti \& Heckman 1999]{calzetti1999}
Calzetti, D. \& Heckman, T. M. 1999, \apj, 519, 27

\bibitem[Coleman, Wu, \& Weedman 1980]{cww1980}
Coleman, G. D., Wu, C., \& Weedman, D. W. 1980, \apjs, 43, 393

\bibitem[Connolly et al. 1995]{connolly1995}
Connolly, A. J., Csabai, I., Szalay, A. S., Koo, D. C., Kron, R. G., \& Munn, J. A. 1995, \aj, 110, 6

\bibitem[Connolly et al. 1997]{connolly1997}
Connolly, A. J., Szalay, A. S., Dickinson, M., SubbaRao, M. U., \& Brunner, R. J. 1997, \apj, 486, 11L

\bibitem[Cowie et al. 1996]{cowie1996}
Cowie, L. L., Songaila, A., Hu, E. M., \& Cohen, J. G. 1996, \apj, 112, 3

\bibitem[Cohen et al. 2000]{cohen2000}
Cohen. J. G. et al. 2000, \apj\ submitted, http://www.ifa.hawaii.edu/\~{}cowie/tts/tts.html

\bibitem[Csabai et al. 1999]{csabai1999} 
Csabai, I., Connolly, A. J., Szalay, A. S., \& Budav\'{a}ri, T. 1999,
 http://tarkus.pha.jhu.edu/\~{}csabai/template/

\bibitem[Dickinson et al. 1999]{dickinson1999}
Dickinson, M. E. et al. 1999 in preparation, http://www.stsci.edu/ftp/science/hdf/clearinghouse/irim/irim\_hdf.html

\bibitem[Ellis et al. 1996]{ellis1996}
Ellis R. S., Colless, M., Broadhurst, T., Heyl, J., \& Glazebrook, K. 1996, \mnras, 280, 235

\bibitem[Fern\'{a}ndez-Soto 1999]{soto_web} Fern\'{a}ndez-Soto, A. http://bat.phys.unsw.edu.au/\~{}fsoto/hdfcat.html

\bibitem[Fern\'{a}ndez-Soto, Lanzetta, \& Yahil 1999]{soto1999} Fern\'{a}ndez-Soto, A., Lanzetta, K.M., \& Yahil,
 A. 1999, \apj, 513, 34

\bibitem[Fitzpatrick 1985]{fitzpatrick1985}
Fitzpatrick, E. L. 1985, \apj, 299, 219

\bibitem[Fitzpatrick 1986]{fitzpatrick1986}
Fitzpatrick, E. L. 1986, \aj, 92, 1068

\bibitem[Fitzpatrick 1999]{fitzpatrick1999}
Fitzpatrick, E. L. 1999, PASP, 111, 63

\bibitem[Giallongo et al. 1998]{giallongo1998}
Giallongo, E., D'odorico, S., Fontana, A., Cristiani, S., Egami, E., Hu,
 E., \& McMahon, R. G. 1998, \aj, 115, 2169 

\bibitem[Gordon, Calzetti, \& Witt 1997]{gordon1997}
Gordon, K. D., Calzetti, D., \& Witt, A. N. 1997, \apj, 487, 625

\bibitem[Gwyn \& Hartwick 1996]{gwyn1996}
Gwyn, S.D.J. \& Hartwick, F.D.A. 1996, \apj, 468, 77L

\bibitem[Hogg et al. 1998]{hogg1998}
Hogg, D.W., Cohen, J.G., Blandford, R., Gwyn, S.D.J., Hartwick, F.D.A., Mobasher, B., Mazzei, P., Sawicki, M.J., Lin, H., Yee, H.K.C., Connolly, A.J., Brunner, R.J., Csabai, I., Dickinson, M., SubbaRao, M.U., Szalay, A.S., Fern\'{a}ndez-Soto, A., Lanzetta, K.M., \& Yahil, A. 1998, \aj, 115, 1418

\bibitem[Kinney, Bohlin, \& Calzetti 1993]{kinney1993}
Kinney, A. L., Bohlin, R. C., \& Calzetti, D. 1993, \apjs, 86, 5

\bibitem[Kinney et al. 1996]{kinney1996}
Kinney, A.L., Calzetti, D., Bohlin, R.C., \& McQuade, K. 1996, \apj, 467,
 38

\bibitem[Kodama \& Arimoto 1997]{ka97}
Kodama, T. \& Arimoto, N. 1997, A\&A, 320, 41

\bibitem[Kodama, Bell, \& Bower 1999]{kodama1999}
Kodama, T., Bell, E. F., \& Bower, R. G. 1999, \mnras, 302, 152

\bibitem[Koo 1985]{koo1985}
Koo, D. C. 1985, \aj, 90, 418

\bibitem[Lanzetta, Yahil, \& Fern\'{a}ndez-Soto 1996]{lanzetta1996}
Lanzetta, K. M., Yahil, A., \& Fern\'{a}ndez-Soto, A. 1996, Nature, 381,
 759

\bibitem[Lanzetta, Yahil, \& Fern\'{a}ndez-Soto]{lanzetta1998}
Lanzetta, K. M., Yahil, A., \& Fern\'{a}ndez-Soto, A. 1998, \aj, 116,
 1066

\bibitem[Lilly et al. 1995]{lilly1995}
Lilly, S. J., Tresse, L., Hammer, F., Crampton, D., \& F\`{e}vre,
 O. 1995, \apj, 455, 108

\bibitem[Loh \& Spillar 1986]{loh1986}
Loh, E. D. \& Spillar, E. J. 1986, \apj, 303, 154

\bibitem[Lowenthal et al. 1997]{lowenthal1997}
Lowenthal, J. D., Koo, D. C., Guzm\'{a}n, R., Gallego, J., Phillips, A. C., Faber, S. M., Vogt, N. P., Illingworth, G. D., \& Gronwall, C. 1997, \apj, 481, 673

\bibitem[Madau 1995]{madau1995} Madau, P. 1995, \apj, 441, 18

\bibitem[Mobasher et al. 1996]{mobasher1996}
Mobasher, B., Rowan-Robinson, M., Georgakakis, A., \& Eaton, N. 1996,
 \mnras, 282, L7

\bibitem[Pr\'{e}vot et al. 1984]{prevot1984}
Pr\'{e}vot, M. L., Lequeux, J., Maurice, E., Pr\'{e}vot, L., \&
 Rocca-Volmerange, B. 1984, A\&A, 132, 389

\bibitem[Sawicki, Lin \& Yee 1997]{sawicki1997}
Sawicki, M. J., Lin, H., \& Yee, H. K. C. 1997, \aj, 113, 1

\bibitem[Sawicki \& Yee 1998]{sawicki1998}
Sawicki, M. J. \& Yee, H. K. C. 1998, \aj, 115, 1329

\bibitem[Sawicki 1999]{sawicki_web}
Sawicki, M. J. http://www.astro.utoronto.ca/\~{}sawicki/

\bibitem[Seaton 1979]{seaton1979} Seaton, M. J. 1979, \mnras, 187, 73

\bibitem[Steidel et al. 1996]{steidel1996}
Steidel, C. C., Giavalisco, M., Dickinson, M., \& Adelberger, K. 1996, \aj, 112, 352

\bibitem[Wang et al. 1998]{wang1998}
Wang, Bahcall, \& Turner 1998, \apj, 116, 2081

\bibitem[Wang et al. 1999]{wang1999}
Wang, Turner, \& Bahcall 1999, astro-ph/9906256

\bibitem[Williams et al. 1996]{williams1996}
Williams, R. E., Blacker, B., Dickinson, M., Dxon, V. D., Ferguson, H. C., Fruchter, A. S., Giavalisco, M., Gililand, R. L., Heyer, I., Katsanis, R., Levay, Z., Lucas, R. A., McElroy, D. B., Petro, L., \& Postman, M. 1996, \aj, 112, 4

\bibitem[Yee et al. 1996]{yee1996}
Yee, H. K. C., Ellingson, E., Bechtold, J., Carlberg, R. G., \&
 Cuillandre, J. C. 1996, \aj, 111, 1783

\bibitem[Yee 1998]{yee1998}
Yee, H. K. C. 1998, astro-ph/9809347

\bibitem[Zepf, Moustakas, \& Davis]{zepf1997} Zepf, S. E., Moustakas,
 L. A., \& Davis, M. 1997, \apj, 474, 1L


\end{thebibliography}
\end{document}